\documentclass{IEEEtran}

\usepackage{color,array,amsthm}
\usepackage{graphicx}

\usepackage{hyperref}
\usepackage{booktabs}
\usepackage{float}
\usepackage{caption}
\usepackage[font=footnotesize, labelfont=it, textfont=it]{subfig}
\usepackage{cite}
\usepackage{amsmath,amssymb,amsfonts}
\usepackage{textcomp}
\usepackage{multirow}
\usepackage{bm}
\newtheorem{assumption}{Assumption}
\usepackage{algorithm}
\usepackage{algpseudocode}  
\usepackage{caption}
\usepackage{siunitx}
\usepackage{makecell}
\usepackage{adjustbox}
\usepackage{tabularx}
\usepackage{orcidlink}

\newcommand{\leftsuper}[3]{{}^{#1}\!{{#2}}^{#3}}
\newlength{\figW}
\makeatletter
\renewcommand\subsubsection{\@startsection{subsubsection}{3}{\z@}%
  {0.5\baselineskip}{0.5\baselineskip}{\normalfont\normalsize\itshape}}
\makeatother

\begin{document}

\title{\bfseries Nonlinear Model Predictive Control \\
for Guidance Law with Target Input Estimation}

\author{
Minho Jang\textsuperscript{1}\orcidlink{0009-0006-7898-7820},
Minjeong Kim\textsuperscript{2}\orcidlink{0000-0002-5374-9360},
and Sungsu Park\textsuperscript{1}\orcidlink{0000-0001-7843-9716}%
\thanks{\textsuperscript{1}\,Minho Jang and Sungsu Park are with the Department of Aerospace Engineering,
Sejong University, Seoul 05006, Republic of Korea.
E-mails: \href{mailto:minho.jang.official@gmail.com}{minho.jang.official@gmail.com}
and \href{mailto:sungsu@sejong.ac.kr}{sungsu@sejong.ac.kr}.}%
\thanks{\textsuperscript{2}\,Minjeong Kim is with the Department of Unmanned Aerial Vehicle Engineering,
Kyungwoon University, Gumi 39160, Republic of Korea.
E-mail: \href{mailto:mjkimrl@gmail.com}{mjkimrl@gmail.com}.}%
}

\IEEEaftertitletext{%
    \vspace{-1.6\baselineskip}
    \centering
    \normalsize\itshape
    (arXiv preprint)\\
    \vspace{0.6\baselineskip}
}

\maketitle

\begin{abstract}
This paper presents a look angle-based nonlinear model predictive control guidance (MPCG) method for missiles equipped with strapdown seekers. Conventional proportional navigation guidance (PNG) requires line-of-sight (LOS) rate measurements, which are not directly available in strapdown systems. 

MPCG instead employs look angles and their derivatives as state variables, eliminating body-rate coupling and associated parasitic feedback. The guidance problem is formulated as a continuous-time optimal control problem (OCP), discretized via the Legendre-Gauss-Radau pseudo-spectral method (LGRPM), and solved as a nonlinear program (NLP) incorporating explicit field-of-view (FOV) and acceleration constraints. Target acceleration at the first step of the prediction horizon is estimated using an adaptive extended Kalman filter (AEKF) integrated with an interacting multiple model (IMM) framework.

Simulation results under single-maneuver scenarios, which include pitch and yaw plane weaving as well as barrel-roll maneuvers, demonstrate that MPCG achieves reliable interception while satisfying operational constraints, outperforming pure PNG (PPNG) in stability and resilience. This indicates that MPCG offers a practical and effective solution for modern missile guidance systems constrained by seeker measurement limitations. 
\end{abstract}


\begingroup
\renewcommand\thefootnote{}
\footnotetext{``This work has been submitted to the IEEE for possible publication. Copyright may be transferred without notice, after which this version may no longer be accessible.''}
\endgroup

\newpage
\newcolumntype{Y}{>{\raggedright\arraybackslash}X} 

\begin{center}
    \textbf{Nomenclature}
\end{center}

\renewcommand{\arraystretch}{0.9}
\scriptsize

\begin{tabularx}{\linewidth}{c X c}
\toprule
\textbf{Symbol} & \centering\textbf{Description} & \textbf{Units} \\
\midrule
$\phi,\, \theta,\, \psi$ & Roll, pitch, yaw angles & rad \\
$p,\, q,\, r$ & Roll, pitch, yaw rates & rad/s \\
$\mu,\, \gamma,\, \chi$ & Bank, flight path, heading angles & rad \\
$\dot{\mu},\, \dot{\gamma},\, \dot{\chi}$ & Bank, flight path, heading angle rates & rad/s \\
$\lambda_{\theta},\, \lambda_{\psi}$ & Pitch, yaw LOS angles & rad \\
$\dot{\lambda}_{\theta},\, \dot{\lambda}_{\psi}$ & Pitch, yaw LOS rates & rad/s \\
$\sigma_{\phi},\, \sigma_{\theta},\, \sigma_{\psi}$ & Roll, pitch, yaw look angles & rad \\
$\dot{\sigma}_{\phi},\, \dot{\sigma}_{\theta},\, \dot{\sigma}_{\psi}$ & Roll, pitch, yaw look angle rates & rad/s \\
${\Omega}^b_{\theta},\, {\Omega}^b_{\psi}$ & Pitch, yaw LOS angular velocities expressed in the body-fixed frame & rad/s \\
$\tilde{{\Omega}}^b_{\theta},\, \tilde{{\Omega}}^b_{\psi}$ & Simplified pitch, yaw LOS angular velocities expressed in the body-fixed frame & rad/s \\
$\bm{Q},\, \bm{Q}_f,\, \bm{R}$ & Weighting matrices of MPC & -- \\
$Q_k,\, R_k$ & Noise covariance matrices of Kalman filter & -- \\
$\hat{x}_{k|k-1},\, \hat{x}_{k|k}$ & Priori, posteriori estimates & -- \\
AEKF & Adaptive Extended Kalman Filter & -- \\
BLOS & Body Line of Sight & -- \\
IMM & Interacting Multiple Model & -- \\
LGRPM & Legendre-Gauss-Radau Pseudo-spectral Method & -- \\
MD & Miss Distance & -- \\
MPCG & Nonlinear Model Predictive Control for Guidance law (\textbf{\textit{proposed}}) & -- \\
PPNG & Pure Proportional Navigation Guidance Law & -- \\
TTI & Time-to-Interception & -- \\
\bottomrule
\end{tabularx}

\renewcommand{\arraystretch}{1.0}
\normalsize

\nocite{*}

\section{Introduction}

\IEEEPARstart{M}{issile} seekers are generally classified into gimbal and strapdown types according to their mechanical structure. A Gimbal seeker, equipped with internal gyros and actuators, mechanically tracks the line-of-sight (LOS) and provide direct and accurate measurements of the LOS angles and LOS rates, offering a wide field-of-view (FOV) \cite{c01}. This capability forms the basis of the proportional navigation guidance (PNG) law \cite{c03}, which uses LOS rates, velocity vector, and a navigation constant to generate acceleration commands. Over decades, PNG has proven effective across various weapon systems, and numerous variants have emerged, including True PNG (TPNG) for exo-atmospheric use \cite{c04}, Pure PNG (PPNG) for endo-atmospheric use \cite{c05}, and Augmented PNG (APNG) incorporating estimated target acceleration \cite{c06, c07}.

\begin{figure}[!htbp]
    \centering
    \subfloat[Gimbal Seeker]{%
        \includegraphics[width=0.55\columnwidth]{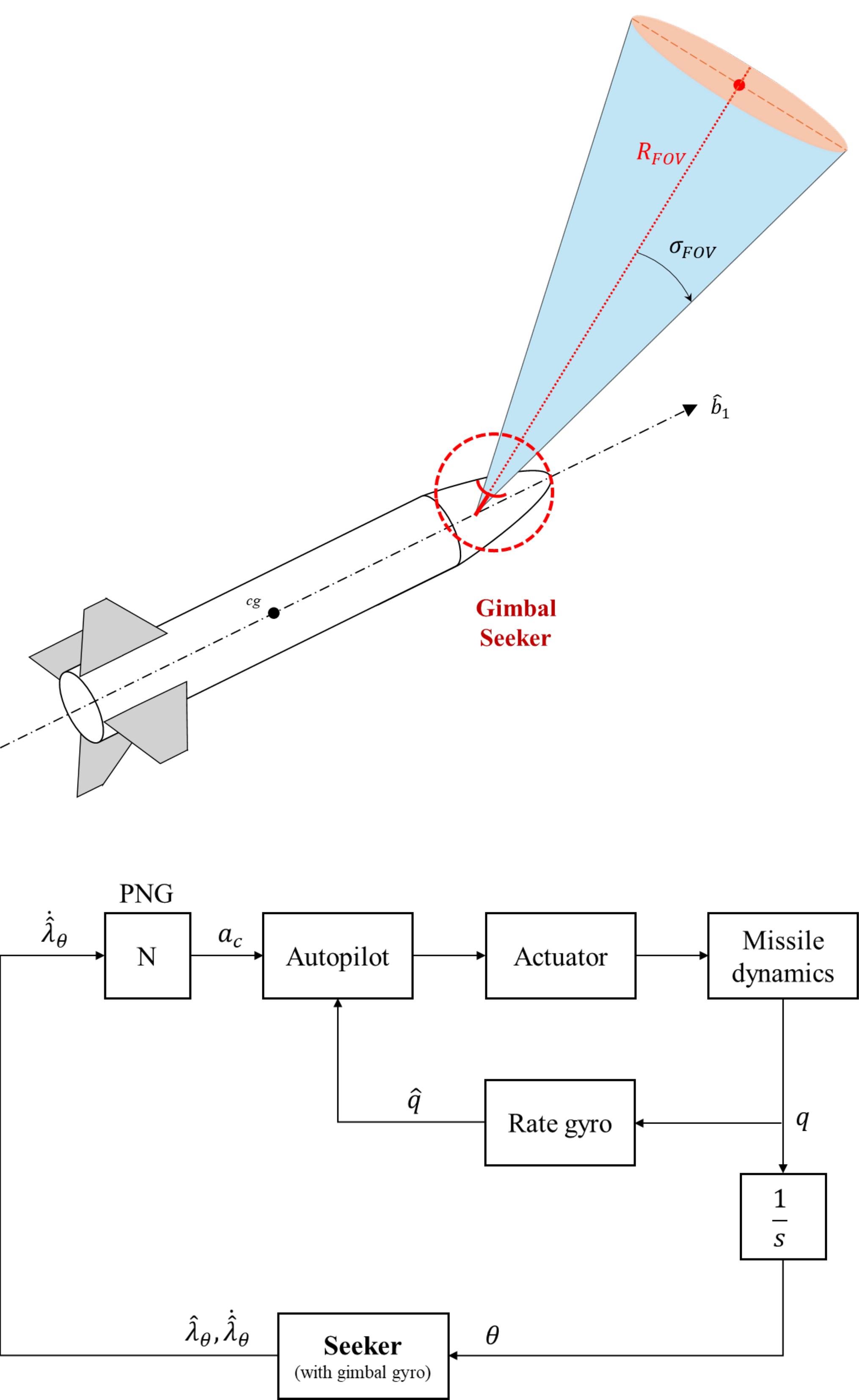}%
        \label{fig:1_gimbal_seeker}
    } \\
    \vfil
    \subfloat[Strapdown Seeker]{%
        \includegraphics[width=0.6\columnwidth]{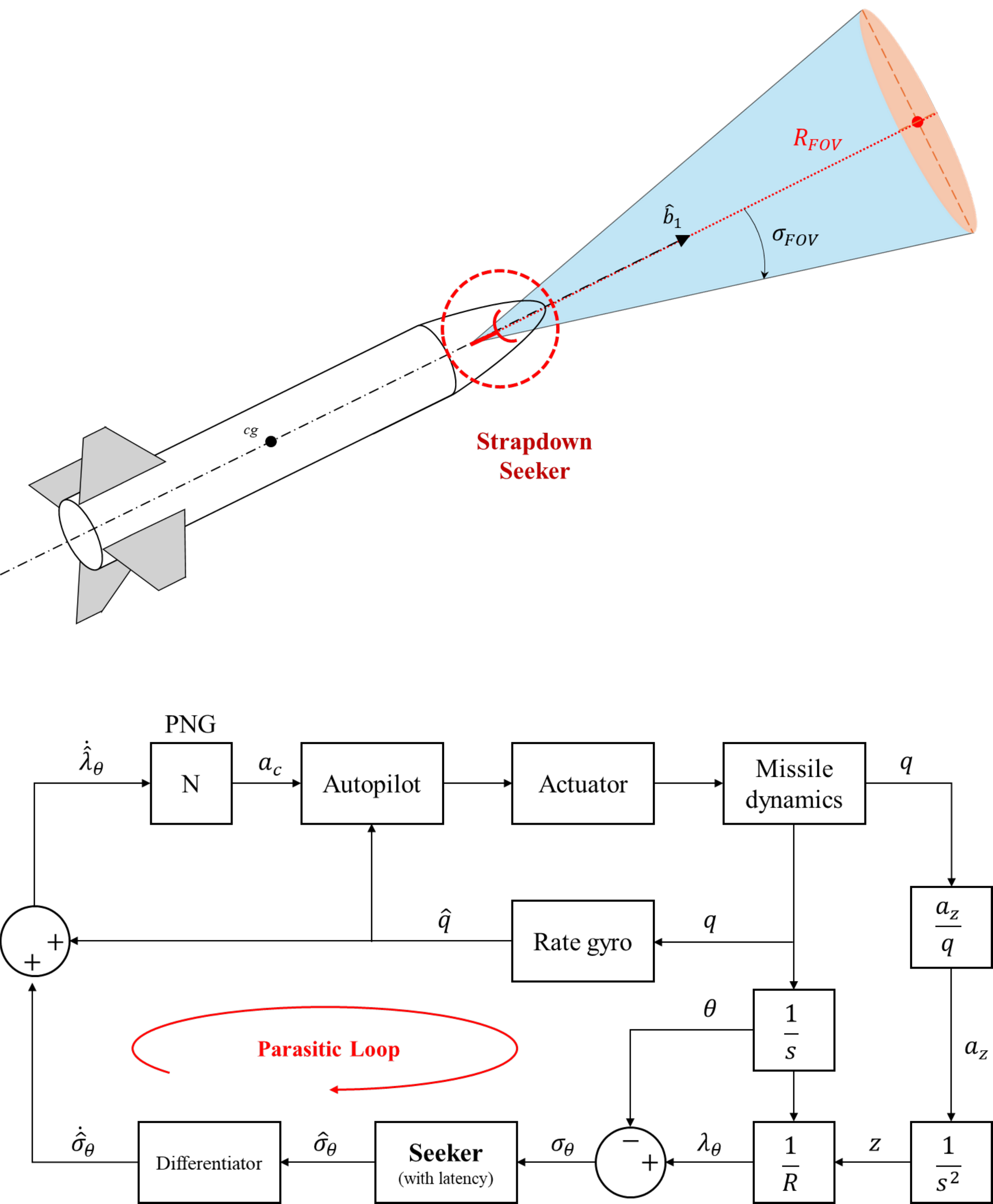}%
        \label{fig:1_strapdown_seeker}
    }
    \vspace{0.1ex}
    \caption[Comparison of seekers]{Comparison of gimbal and strapdown seekers \\
    \footnotesize
    A gimbal seeker provides a wide FOV and directly measures the LOS rates via gimbal rotation, whereas a strapdown seeker estimates it indirectly from body rotation, making it more vulnerable to parasitic loop effects caused by estimation delay and phase mismatch.}
    \label{fig:1_1_comparison_seekers}
\end{figure}

\begin{figure*}[!htbp]
  \centering
  \includegraphics[width=\textwidth]{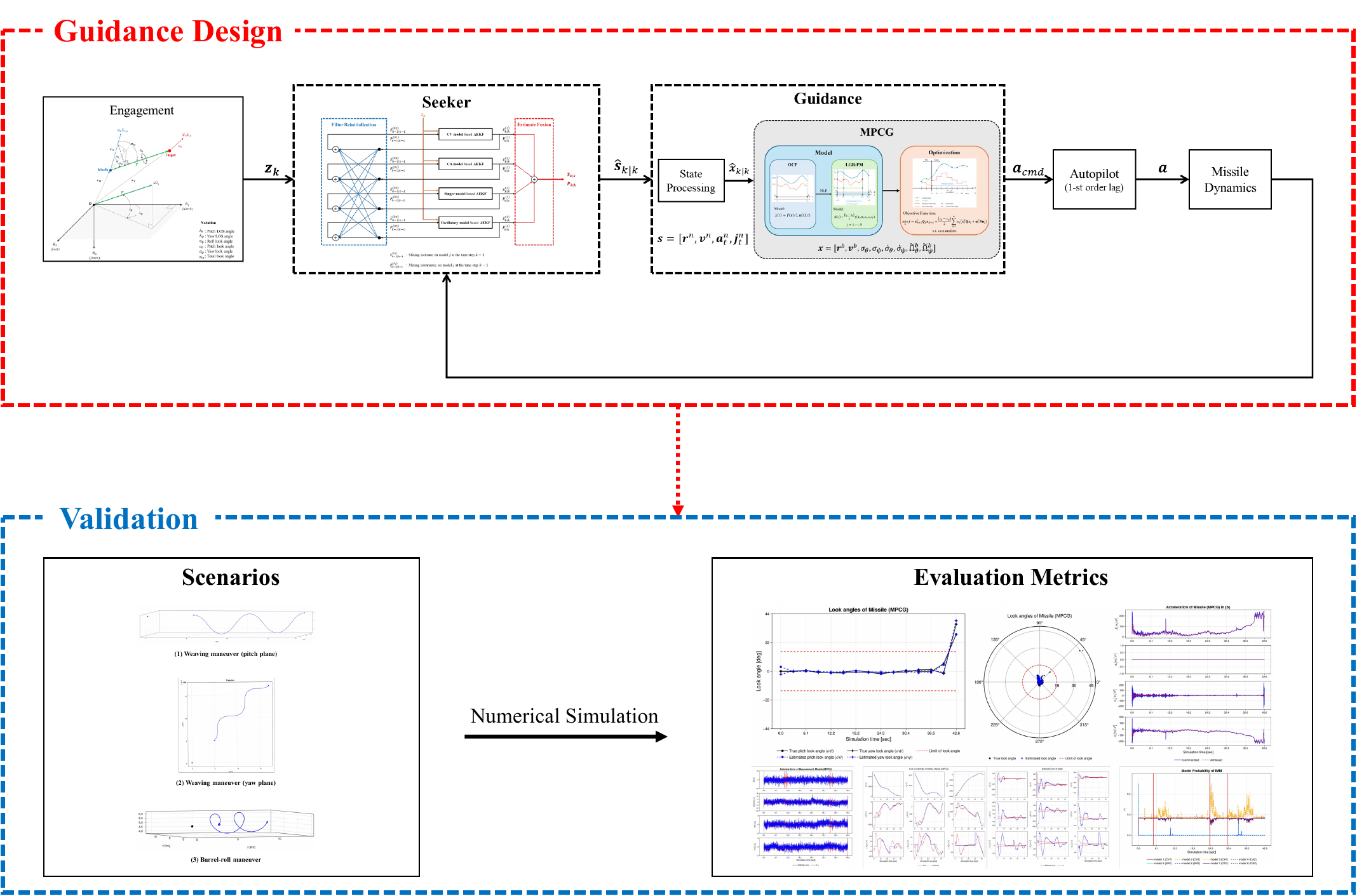}
  \caption{Architecture of the guidance design and validation}
  \label{fig:1_2_overview_architecture}
\end{figure*}

The use of strapdown seekers, as in the PAC-3 missile, has grown due to reduced weight, simplicity, and cost. However, without gimbals, they cannot directly measure LOS rates, instead providing look angles between the missile body axis and the target. For PNG, look angle rates are typically obtained by differentiating measured look angles, then combined with body rates to reconstruct LOS rates. This approach suffers from phase mismatch: look angle measurements incur latency due to signal processing, while body rates are real-time, producing parasitic loops that degrade Guidance, Navigation, and Control (GNC) performance. Prior work has proposed delay-compensated estimation methods, including prediction-based corrections and Kalman filter variants \cite{c08, c09, c10, c11, c12}.

Several studies have proposed guidance laws directly utilizing look angle measurements to accommodate strapdown seeker constraints \cite{c13, c13_0, c13_1, c13_2}. 
These methods avoid explicit LOS rate reconstruction, thereby mitigating latency-induced feedback issues, but often employ relatively simple control schemes without leveraging advanced predictive optimization. Other research has explored model predictive control guidance (MPCG) formulations for missile interception \cite{c16, c21}. 
While these approaches benefit from the constraint-handling and nonlinear modeling capabilities of MPC, they generally assume constant or known target acceleration, lacking an online estimation mechanism.

Despite these advances, PNG and existing MPCG variants remain limited in addressing both the measurement constraints of strapdown seekers and the uncertainties of maneuvering targets. This paper proposes a look angle-based nonlinear model predictive control guidance (MPCG) method that integrates online target acceleration estimation via an Adaptive Extended Kalman Filter (AEKF) with an Interacting Multiple Model (IMM) framework. By combining look angle-based state representation with constraint-aware predictive optimization and maneuver-adaptive estimation, the proposed architecture ensures accurate and reliable guidance performance against agile targets.

The remainder of this paper is organized as follows. Section \ref{sec:2_guidance_algorithm} describes the proposed guidance algorithm, presenting the equations of motion (EOM), engagement geometry, OCP formulation, and the mathematical structure of the MPCG. 

Section \ref{sec:3_target_estimation} focuses on the target input estimation framework, detailing the architecture of the AEKF–IMM and its handling of various target maneuvers. 

Section \ref{sec:4_numerical} provides numerical simulation results, evaluating performance metrics such as miss distance (MD) and time-to-interception (TTI) while verifying compliance with FOV constraints. 

Finally, Section \ref{sec:5_conclusion} concludes the paper and discusses potential directions for future research.

\subsection{Research Contributions and Overview}

The motivation and contributions of this work are as follows.

\textbf{Motivation:}

Strapdown seekers measure look angles rather than LOS rates, creating limitations for conventional guidance laws such as PNG. This study proposes a guidance law tailored for strapdown seekers, directly utilizing look angle dynamics to overcome these constraints.

\textbf{Key Contributions:}

\begin{itemize}
    \item Developed MPCG, a look angle-based guidance scheme optimized for strapdown seeker limitations.

    \item Formulated a constraint-aware NMPC by transcribing the nonlinear OCP via LGRPM.

    \item Designed an AEKF–IMM estimator capable of handling diverse target maneuvers with adaptive noise adjustment.

    \item Validated MPCG in multiple engagement scenarios, achieving reliable terminal guidance performance.
\end{itemize}

\textbf{Overview:}

The proposed architecture in Fig. \ref{fig:1_2_overview_architecture} integrates a strapdown seeker, an IMM-based estimator, and an NMPC guidance law. The seeker measures look angles and range, which are processed by the AEKF–IMM to estimate target acceleration across multiple motion models (CV, CA, Singer, Oscillatory). The estimated acceleration is incorporated into the NMPC prediction model, discretized with LGRPM, to generate optimal acceleration commands under FOV and maneuver constraints.

Performance was evaluated in three representative scenarios—pitch-plane weaving, yaw-plane weaving, and barrel-roll—using metrics such as final MD, TTI, constraint satisfaction, and estimation accuracy. Results confirmed MPCG’s stability and adaptability against agile targets.


\section{Guidance Algorithm} \label{sec:2_guidance_algorithm}


\subsection{Notation for the Equations}

A vector $\bm{v}$ expressed in coordinate systems $\{a\}$ and $\{b\}$ is written as

\begin{equation}
\begin{aligned}
    \bm{v}^{a} &= v_{a_1} \hat{a}_1 + v_{a_2} \hat{a}_2 + v_{a_3} \hat{a}_3 \\
    \bm{v}^{b} &= v_{b_1} \hat{b}_1 + v_{b_2} \hat{b}_2 + v_{b_3} \hat{b}_3
\end{aligned}  
\end{equation}

or in column form $\bm{v}^{a} = \left[ v_{a_1},\, v_{a_2},\, v_{a_3}  \right]^\top$, $\bm{v}^b = \left[ v_{b_1},\, v_{b_2},\, v_{b_3}  \right]^\top$. Here, $\hat{a}$ and $\hat{b}$ are the unit coordinate vectors of frames $\{a\}$ and $\{b\}$, respectively.

The Direction Cosine Matrix (DCM) $C^a_b$ transforms a vector $\bm{v}$ from frame $\{a\}$ to frame $\{b\}$ and is defined as $\left( C^a_b \right)_{ij} = \hat{a}_i \cdot \hat{b}_j$. Also, DCM has the following properties.

\begin{equation} \label{eq:2_DCM_property}
        \bm{v}^a = C^a_b \bm{v}^b,\quad
        \left(C^a_b\right)^{-1} = \left(C^a_b\right)^{\mathrm{T}},\quad
        C^a_c = C^a_b C^b_c
\end{equation}

\begin{figure}[!htbp]
  \centering
  \includegraphics[width=0.9\columnwidth]{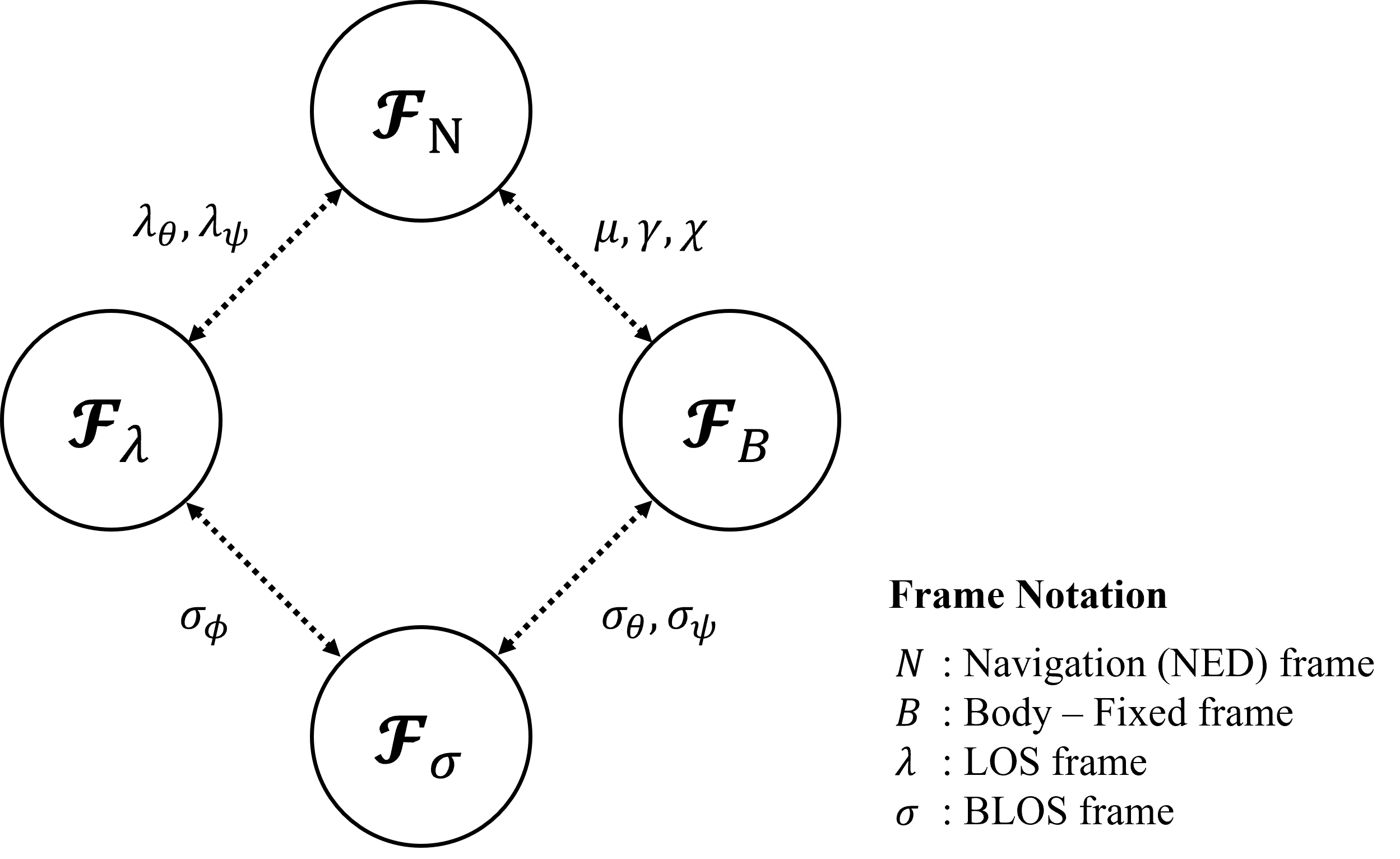}
  \caption{Relationships among coordinate systems of the missile}
  \label{fig:2_1_frames}
\end{figure}

In this paper, four coordinate frames, as shown in Fig. \ref{fig:2_1_frames}, are used: navigation frame $\{n\}$, body-fixed frame $\{b\}$, LOS frame $\{\lambda\}$, and body LOS (BLOS) frame $\{\sigma\}$. The transformations are parameterized by the bank angle $\mu$, flight path angle $\gamma$, and heading angle $\chi$ between $\{n\}$ and $\{b\}$. The look angles $(\sigma_\theta, \sigma_\phi)$ describe the transformation from $\{b\}$ to $\{\sigma\}$, while the bank angle of the look angle $\sigma_\phi$ connects $\{\sigma\}$ to $\{\lambda\}$. The elevation and azimuth of the LOS vector $(\lambda_\theta, \lambda_\phi)$ are defined with respect to $\{n\}$.

\subsection{Equation of Motion}

\begin{figure}[!htbp]
  \centering
  \includegraphics[width=\columnwidth]{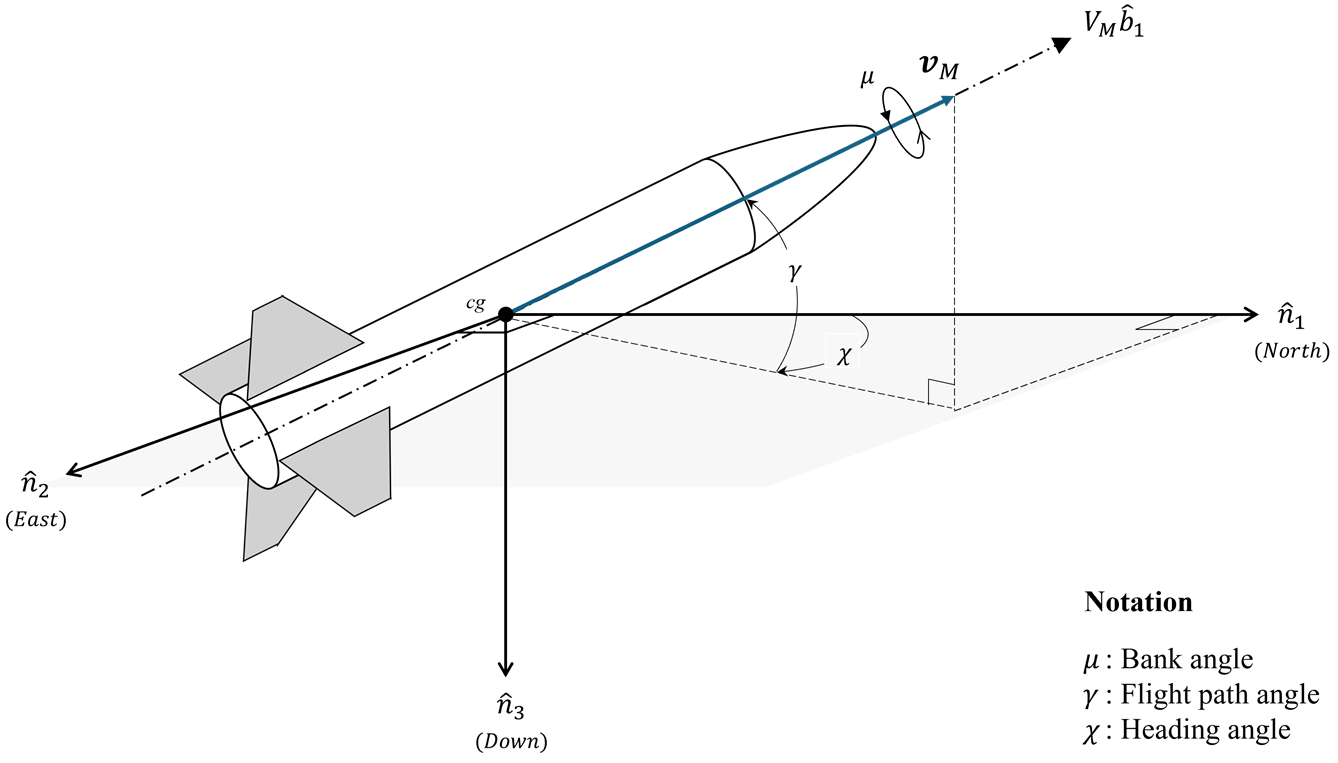}
  \caption{Relationship of navigation frame to body-fixed frame}
  \label{fig:2_2_navigation}
\end{figure}

The missile and the target were modeled using three-degree-of-freedom (3-DOF) point-mass equations of motion, which are formulated in (\ref{eq:2_eom}). The Fig. \ref{fig:2_2_navigation} illustrates the navigation frame $\{n\}$, the flight path frame $\{f\}$, and the body-fixed frame $\{b\}$ of the missile.

The navigation frame adopts the North-East-Down (NED) convention, with the z-axis pointing downward and its origin fixed on the Earth's surface, providing a stationary inertial reference for motion differentiation.

The missile velocity in the navigation frame is used to compute the flight path angle $(\gamma)$ and heading angle $(\chi)$. Together with the DCM, these define the transformation from the frame $\{n\}$ to the frame $\{f\}$ via 3–2 Euler angle rotations.

{\footnotesize
\begin{align}
    \dot{\bm{r}}^n_M &= \bm{v}^n_M, \quad
    \dot{\bm{v}}^n_M = \bm{a}^n_M \label{eq:2_eom} \\
        \bm{v}^n_M &= v^n_{1,m} \hat{n}_1 + v^n_{2,m} \hat{n}_2 + v^n_{3,m} \hat{n}_3,\quad V_M = \left\| \bm{v}^n_M \right\| \\
        C^n_f &= C(z, \chi) \, C(y, \gamma) \label{eq:2_DCM_nf} \\
        \gamma &= \operatorname{atan} \left( \frac{-v^n_{3,m}}{\sqrt{(v^n_{1,m})^2 + (v^n_{2,m})^2}} \right), 
        \; -\frac{\pi}{2} < \gamma < \frac{\pi}{2} \label{eq:2_fpa} \\
        \chi &= \operatorname{atan2} \left( \frac{v^n_{2,m}}{v^n_{1,m}} \right), 
        \; -\pi < \chi \leq \pi \nonumber
\end{align}
}

The flight path angle $\gamma$ is the elevation of the missile’s velocity from the horizontal plane, and the heading angle $\chi$ is its azimuth relative to the $x$-axis of the navigation frame. The flight path frame is transformed to the body-fixed frame by a rotation about its $x$-axis by the bank angle $\mu$, representing the missile’s roll orientation.

{\footnotesize
\begin{align}
    \bm{v}_M^f &= V_M \hat{\bm{f}}_1, \quad 
    \bm{v}_M^b = V_M \hat{\bm{b}}_1, \quad 
    \hat{\bm{f}}_1 = \hat{\bm{b}}_1 \\
    \bm{a}_M^f &= a_{1,m}^f \hat{\bm{f}}_1 + a_{2,m}^f \hat{\bm{f}}_2 + a_{3,m}^f \hat{\bm{f}}_3 \nonumber \\
    C^f_b &= C(x, \mu) \label{eq:2_DCM_fb} \\
    \mu &= \operatorname{atan2} \left( \frac{a_{2,m}^f}{-a_{3,m}^f} \right), \quad -\pi < \mu \leq \pi
\end{align}
}

Consequently, using (\ref{eq:2_DCM_nf}) and (\ref{eq:2_DCM_fb}), the DCM from $\{n\}$ to $\{b\}$ and the angular velocity of $\{b\}$ with respect to $\{n\}$ are obtained as:

{\footnotesize
\begin{align}
    C_b^n &= C_f^n C_b^f = C(z,\chi) C(y,\gamma) C(x,\mu)  \label{eq:2_DCM_nb}  \\
    &= 
    \begin{bmatrix}
        c_\chi & -s_\chi & 0 \\
        s_\chi &  c_\chi & 0 \\
        0      & 0       & 1
    \end{bmatrix}
    \begin{bmatrix}
        c_\gamma & 0 & s_\gamma \\
        0        & 1 & 0        \\
        -s_\gamma& 0 & c_\gamma
    \end{bmatrix}
    \begin{bmatrix}
        1  & 0    & 0     \\
        0  & c_\mu & -s_\mu \\
        0  & s_\mu & c_\mu
    \end{bmatrix} \nonumber   \\
    &=
    \begin{bmatrix}
        c_\chi c_\gamma & -s_\chi c_\mu + c_\chi s_\gamma s_\mu & s_\chi s_\mu + c_\chi s_\gamma c_\mu \\
        s_\chi c_\gamma & c_\chi c_\mu + s_\chi s_\gamma s_\mu & -c_\chi s_\mu + s_\chi s_\gamma c_\mu \\
        -s_\gamma       & c_\gamma s_\mu & c_\gamma c_\mu
    \end{bmatrix} \nonumber \\
    \leftsuper{n}{\bm{\omega}}{b}&= \dot{\chi} \hat{n}_3 + \dot{\gamma} \hat{f}_2 + \dot{\mu} \hat{b}_1 \\
    \bm{\omega}_{nb}^b &= 
    \begin{bmatrix}
        p \\
        q \\
        r
    \end{bmatrix}
    = C_n^b 
    \begin{bmatrix}
        0 \\ 0 \\ \dot{\chi}
    \end{bmatrix}
    + C_f^b 
    \begin{bmatrix}
        0 \\ \dot{\gamma} \\ 0
    \end{bmatrix}
    + 
    \begin{bmatrix}
        \dot{\mu} \\ 0 \\ 0
    \end{bmatrix} \label{eq:2_omega_b_nb} \\
    &=
    \begin{bmatrix}
        -\dot{\chi}s_\gamma + \dot{\mu} \\
        \dot{\chi} c_\gamma s_\mu + \dot{\gamma} c_\mu \\
        \dot{\chi} c_\gamma c_\mu - \dot{\gamma} s_\mu
    \end{bmatrix} \nonumber
\end{align}
}

In (\ref{eq:2_omega_b_nb}), $\bm{\omega}_{nb}^b$ is the angular velocity of $\{b\}$ with respect to $\{n\}$ expressed in $\{b\}$, with $p$, $q$, and $r$ denoting roll, pitch, and yaw rates measured by onboard gyros. In the 3-DOF point-mass model, representing $\bm{\omega}_{nb}^b$ via flight path, heading, and bank angles and their derivatives requires the following assumption.

\vspace{1ex}

\begin{assumption}\label{assump:aoa}
The angle of attack ($\alpha$) and the sideslip angle ($\beta$), representing the rotational misalignment between the body-fixed frame and the wind-axis frame, are assumed to be negligibly small.
\end{assumption}

\vspace{1ex}


From (\ref{eq:2_fpa}), the time derivatives of the flight path and heading angles are:

{\footnotesize
\begin{equation}
\label{eq:2_fpa_rates}
\begin{split}
\dot{\gamma} &=
\frac{ v_{3,m}^n ( v_{1,m}^n a_{1,m}^n + v_{2,m}^n a_{2,m}^n )
      - a_{3,m}^n \big[(v_{1,m}^n)^2 + (v_{2,m}^n)^2 \big] }
     { \|\bm{v}_M^n\|^2 \sqrt{(v_{1,m}^n)^2 + (v_{2,m}^n)^2} } \\
\dot{\chi} &=
\frac{ v_{1,m}^n a_{2,m}^n - v_{2,m}^n a_{1,m}^n }
     { (v_{1,m}^n)^2 + (v_{2,m}^n)^2 }
\end{split}
\end{equation}
}

\subsection{Engagement Geometry}

\begin{figure}[!htbp]
  \centering
  \includegraphics[width=\columnwidth]{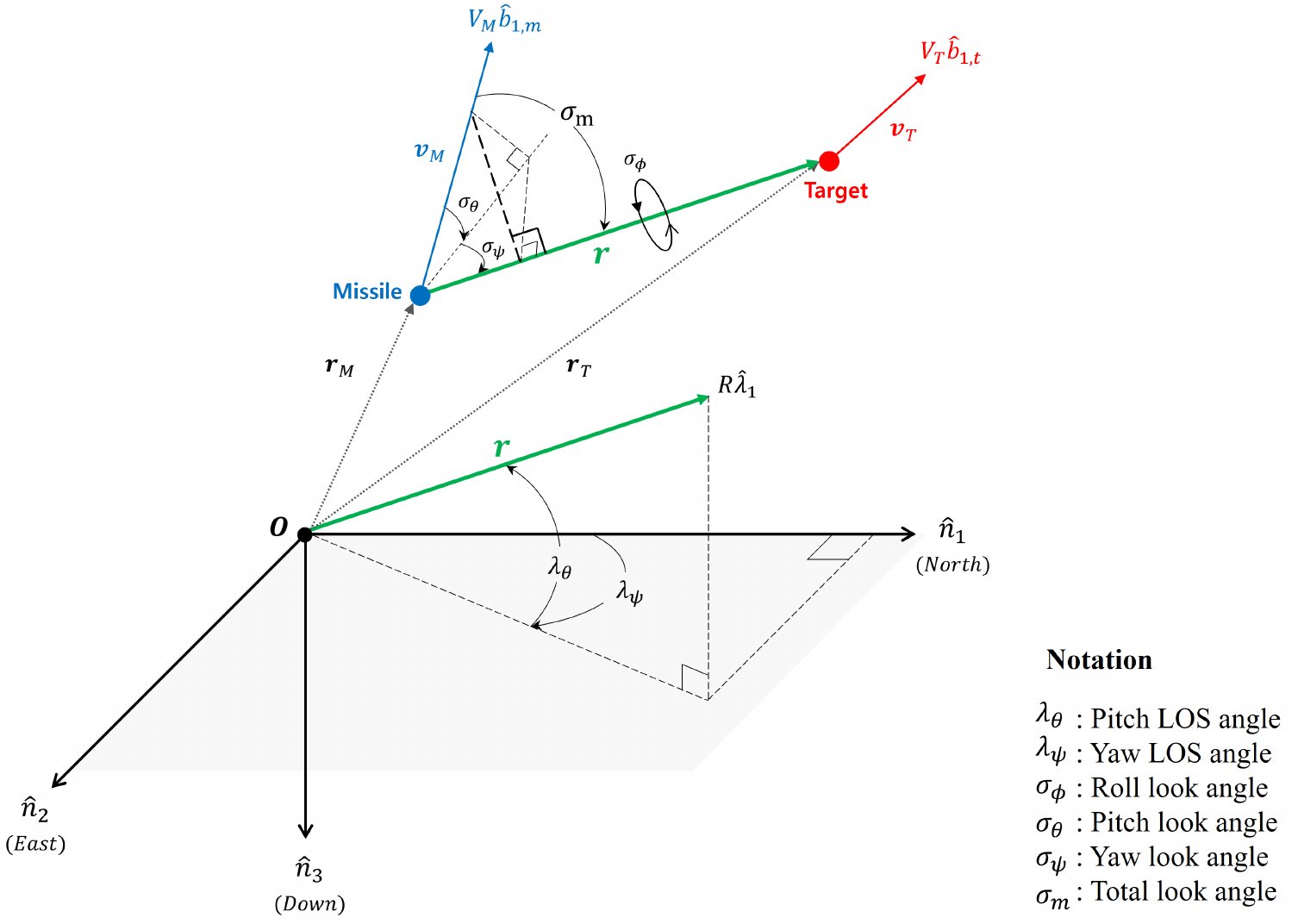}
  \caption{Engagement geometry in three-dimensional space}
  \label{fig:2_3_geometry}
\end{figure}

In Fig. \ref{fig:2_3_geometry}, four coordinate frames are applied, as defined in Fig. \ref{fig:2_1_frames}. $\lambda_\theta$ and $\lambda_\psi$ are the elevation and azimuth of the LOS vector, while, $\sigma_\theta$ and $\sigma_\psi$ are the elevation and azimuth of the look angle, referred to here as pitch/yaw LOS angle and pitch/yaw look angle, respectively. The look angles represent the angular deviation between $\{b\}$ and $\{\lambda\}$, indicating the target’s direction relative to $\hat{b}_1$. The missile–target engagement geometry is formulated in both $\{n\}$ and $\{\lambda\}$ frames as follows:

{\footnotesize
\begin{align}
    \bm{r}^n &= r^n_1 \hat{n}_1 + r^n_2 \hat{n}_2 + r^n_3 \hat{n}_3  \label{eq:2_r_v} \\
        \bm{v}^n &= v^n_1 \hat{n}_1 + v^n_2 \hat{n}_2 + v^n_3 \hat{n}_3 \nonumber \\ 
        \bm{r} &= \bm{r}_T - \bm{r}_M = R \hat{\lambda}_1, \quad
        \bm{r}^\lambda = \begin{bmatrix}
            R  \\ 0 \\ 0
        \end{bmatrix} 
         \label{eq:2_eom_rel} \\ 
        \bm{v} &= \frac{{}^{n}\!d\bm{r}}{dt} = \bm{v}_T - \bm{v}_M \label{eq:2_eom_vel} \\
        &= \frac{\leftsuper{\lambda}{d\bm{r}}{}}{dt} +
        \leftsuper{n}{\bm{\omega}}{\lambda}
        \times \bm{r}
        = \dot{R} \hat{\lambda}_1 + R \leftsuper{n}{\bm{\omega}}{\lambda} \times \hat{\lambda}_1 \nonumber 
        \\ 
        \lambda_\theta &= \operatorname{atan} \left( \frac{-r_3^n}{\sqrt{(r_1^n)^2 + (r_2^n)^2}} \right), 
        \; -\frac{\pi}{2} < \lambda_\theta < \frac{\pi}{2} \label{eq:2_LOS_angle} \\
        \lambda_\psi &= \operatorname{atan2} \left( \frac{r_2^n}{r_1^n} \right), 
        \; -\pi < \lambda_\psi \leq \pi
        \nonumber \\ 
        C^n_\lambda &= C(z, \lambda_\psi) C(y, \lambda_\theta) \label{eq:2_dcm_nlambda} \\
        &= \begin{bmatrix}
        c_{\lambda_\psi} & -s_{\lambda_\psi} & 0 \\
        s_{\lambda_\psi} & c_{\lambda_\psi} & 0 \\
        0 & 0 & 1
        \end{bmatrix} 
        \begin{bmatrix}
        c_{\lambda_\theta} & 0 &  s_{\lambda_\theta} \\
        0 & 1 & 0 \\
        -s_{\lambda_\theta} & 0 &  c_{\lambda_\theta}
        \end{bmatrix} \nonumber \\
        &=  \begin{bmatrix}
        c_{\lambda_\psi}c_{\lambda_\theta} & -s_{\lambda_\psi} &  c_{\lambda_\psi}s_{\lambda_\theta} \\
        s_{\lambda_\psi}c_{\lambda_\theta} & c_{\lambda_\psi} & s_{\lambda_\psi}s_{\lambda_\theta} \\
        -s_{\lambda_\theta} & 0 &  c_{\lambda_\theta}
        \end{bmatrix} \nonumber
\end{align}
}

In (\ref{eq:2_eom_rel}), The $\bm{r}_M$ is a displacement from the origin of navigation frame to missile. The $\bm{r}$ represents the relative position vector between the missile and the target, with magnitude R. The relative velocity $\bm{v}$ as in (\ref{eq:2_eom_vel}) can be obtained by differentiating the relative position vector in the navigation frame, where $\leftsuper{n}{\bm{\omega}}{\lambda}$ is the angular velocity of the LOS frame $\{\lambda\}$ with respect to the navigation frame $\{n\}$. 
Similarly, the angular velocity of the LOS frame with respect to the navigation frame and the relative velocity, expressed in the LOS frame is given as follows.

{\footnotesize
\begin{align}  \label{eq:2_angular_vel_n_los}
    \leftsuper{n}{\bm{\omega}}{\lambda} &= \dot{\lambda}_\psi \hat{n}_3 + \dot{\lambda}_\theta \hat{\lambda}_2 \\
    \bm{\omega}^\lambda_{n\lambda} &= C^\lambda_n \begin{bmatrix}
        0 \\ 0 \\ \dot{\lambda}_\psi
    \end{bmatrix} + 
    \begin{bmatrix}
        0 \\ \dot{\lambda}_\theta \\ 0
    \end{bmatrix} 
    = \begin{bmatrix}
        -\dot{\lambda}_\psi s_{\lambda_\theta} \\ \dot{\lambda}_\theta \\
        \dot{\lambda}_\psi c_{\lambda_\theta}
    \end{bmatrix} \label{eq:2_angular_vel_n_los2}
\end{align}
}

However, the conventional on-board seeker cannot measure the angular velocity $\bm{\omega}^\lambda_{n\lambda}$, which includes the angular rate along $\hat{\lambda}_1$. So, another angular velocity $\bm{\Omega}$, referred to as the LOS angular velocity, is defined as follows.

{\footnotesize
\begin{equation} \label{eq:2_LOS_angular_vel}
    \begin{aligned}
    \bm{\Omega} &= \leftsuper{n}{\bm{\omega}}{\lambda} - \left( {\leftsuper{n}{\bm{\omega}}{\lambda} \cdot \hat{\lambda}_1 } \right) \hat{\lambda}_1
                =  \frac{ \bm{r} \times \bm{v} }{ \bm{r} \cdot \bm{r} } \\
    \bm{\Omega}^\lambda &= \begin{bmatrix}
        0 \\ \dot{\lambda}_\theta \\ \dot{\lambda}_{\psi} c_{\lambda_\theta}
    \end{bmatrix}
    \end{aligned}
\end{equation}
}

In addition, the pitch and yaw LOS rates — the time derivatives of the respective angles — as expressed in equations (\ref{eq:2_angular_vel_n_los}) to (\ref{eq:2_LOS_angular_vel}), are defined through the following derivation process starting from (\ref{eq:2_LOS_angle}).

{\footnotesize
\begin{equation}
    \begin{aligned}
\dot{\lambda}_\theta &= \frac{r_3^n \left( r_1^n v_1^n + r_2^n v_2^n \right) - v_3^n \left[ (r_1^n)^2 + (r_2^n)^2 \right]}{\left[ (r_1^n)^2 + (r_2^n)^2 + (r_3^n)^2 \right] \sqrt{(r_1^n)^2 + (r_2^n)^2}} \\
\dot{\lambda}_\psi &= \frac{r_1^n v_2^n - r_2^n v_1^n}{(r_1^n)^2 + (r_2^n)^2}
    \end{aligned}
\end{equation}
}

The transformation between the body-fixed frame $\{b\}$ and the BLOS frame $\{\sigma\}$ is defined by the pitch and yaw look angles, which specify the orientation of $\{\sigma\}$ relative to $\{b\}$. The BLOS frame, aligned with the target direction from the missile’s $\hat{b}_1$-axis, yields the following DCM:

{\footnotesize
\begin{align}
        \bm{r}^b &= C_n^b \bm{r}^n = r_1^b \, \hat{b}_1 + r_2^b \, \hat{b}_2 + r_3^b \, \hat{b}_3 \label{eq:2_rbvb} \\
    \bm{v}^b &= C_n^b \bm{v}^n = v_1^b \, \hat{b}_1 + v_2^b \, \hat{b}_2 + v_3^b \, \hat{b}_3 \nonumber \\
    \sigma_\theta &= \operatorname{atan}\left( \frac{-r_3^b}{\sqrt{(r_1^b)^2 + (r_2^b)^2}} \right), 
    \; -\frac{\pi}{2} < \sigma_\theta < \frac{\pi}{2} \label{eq:2_look_angle} \\
    \sigma_\psi &= \operatorname{atan2}\left( \frac{r_2^b}{r_1^b} \right), 
    \; -\pi < \sigma_\psi \leq \pi \nonumber \\
    \sigma_m &= \operatorname{acos}\left( \bm{\hat{r}}^b \cdot \bm{\hat{v}}_M^b \right) \approx \operatorname{acos}\left( \cos{\sigma_\theta} \cos{\sigma_\psi} \right) \label{eq:2_total_look_angle} \\
    C_\sigma^b &= C(z, \sigma_\psi) \, C(y, \sigma_\theta) \label{eq:2_DCM_b_look}  \\
        &=
        \begin{bmatrix}
        c_{\sigma_\psi} & -s_{\sigma_\psi} & 0 \\
        s_{\sigma_\psi} & c_{\sigma_\psi} & 0 \\
        0 & 0 & 1
        \end{bmatrix}
        \begin{bmatrix}
        c_{\sigma_\theta} & 0 & s_{\sigma_\theta} \\
        0 & 1 & 0 \\
        -s_{\sigma_\theta} & 0 & c_{\sigma_\theta}
        \end{bmatrix} \nonumber \\
        &=
        \begin{bmatrix}
        c_{\sigma_\psi} c_{\sigma_\theta} & -s_{\sigma_\psi} & c_{\sigma_\psi} s_{\sigma_\theta} \\
        s_{\sigma_\psi} c_{\sigma_\theta} & c_{\sigma_\psi} & s_{\sigma_\psi} s_{\sigma_\theta} \\
        -s_{\sigma_\theta} & 0 & c_{\sigma_\theta}
        \end{bmatrix} \nonumber
\end{align}
}

In (\ref{eq:2_rbvb}), $r^b$ and $v^b$ are the relative position and velocity in the frame $\{b\}$. The total look angle $\sigma_m$, corresponding to the seeker FOV, is obtained from the dot product of their unit vectors and approximated by pitch and yaw look angles under the small-angle assumption as like (\ref{eq:2_total_look_angle}). Using these vectors and look angles, the angular velocity of the BLOS frame with respect to the frame $\{b\}$, and the pitch/yaw look angle rates are given as follows:

{\footnotesize
\begin{align}
\leftsuper{b}{\bm{\omega}}{\sigma} &= \dot{\sigma}_\psi \hat{b}_3 + \dot{\sigma}_\theta \hat{\sigma}_2 \\
\bm{\omega}_{b\sigma}^b &= 
\begin{bmatrix}
0 \\
0 \\
\dot{\sigma}_\psi
\end{bmatrix}
+ C_\sigma^b
\begin{bmatrix}
0 \\
\dot{\sigma}_\theta \\
0
\end{bmatrix}
=
\begin{bmatrix}
-\dot{\sigma}_\theta s_{\sigma_\psi} \\
\dot{\sigma}_\theta c_{\sigma_\psi} \\
\dot{\sigma}_\psi
\end{bmatrix}  \label{eq:2_angular_vel_body_look} \\
\dot{\sigma}_\theta &= \frac{
r_3^b (r_1^b v_1^b + r_2^b v_2^b) - v_3^b \left[ (r_1^b)^2 + (r_2^b)^2 \right]
}{
\left[ (r_1^b)^2 + (r_2^b)^2 + (r_3^b)^2 \right] \sqrt{(r_1^b)^2 + (r_2^b)^2}
} \label{eq:2_look_angle_rates} \\
\dot{\sigma}_\psi &= \frac{
r_1^b v_2^b - r_2^b v_1^b
}{
(r_1^b)^2 + (r_2^b)^2
} \nonumber
\end{align}
}

Finally, the DCM and the angular velocity between the BLOS frame and the LOS frame are derived. The transformation between these two frames is achieved by rotating the BLOS frame around its x-axis $(\hat{\sigma}_1)$ by the roll look angle $(\sigma_\phi)$, and the DCM and the angular velocity are provided below.

{\footnotesize
\begin{align}
    \sigma_\phi &= -\phi \approx -\mu \label{eq:2_roll_look_angle_roll} \\
       C_\lambda^\sigma &= C(x, \sigma_\phi) 
    =
    \begin{bmatrix}
    1 & 0 & 0 \\
    0 & c_{\sigma_\phi} & -s_{\sigma_\phi} \\
    0 & s_{\sigma_\phi} & c_{\sigma_\phi}
    \end{bmatrix}  \label{eq:2_DCM_look_los} \\
    \leftsuper{\sigma}{\bm{\omega}}{\lambda} &= \dot{\sigma}_\phi \hat{\sigma}_1, \qquad 
\bm{\omega}^\sigma_{\sigma\lambda} =
\begin{bmatrix}
\dot{\sigma}_\phi \\
0 \\
0
\end{bmatrix} \label{eq:2_angular_vel_look_los}
\end{align}
}

\begin{figure}[!htbp]
  \centering
  \includegraphics[width=0.5\columnwidth]{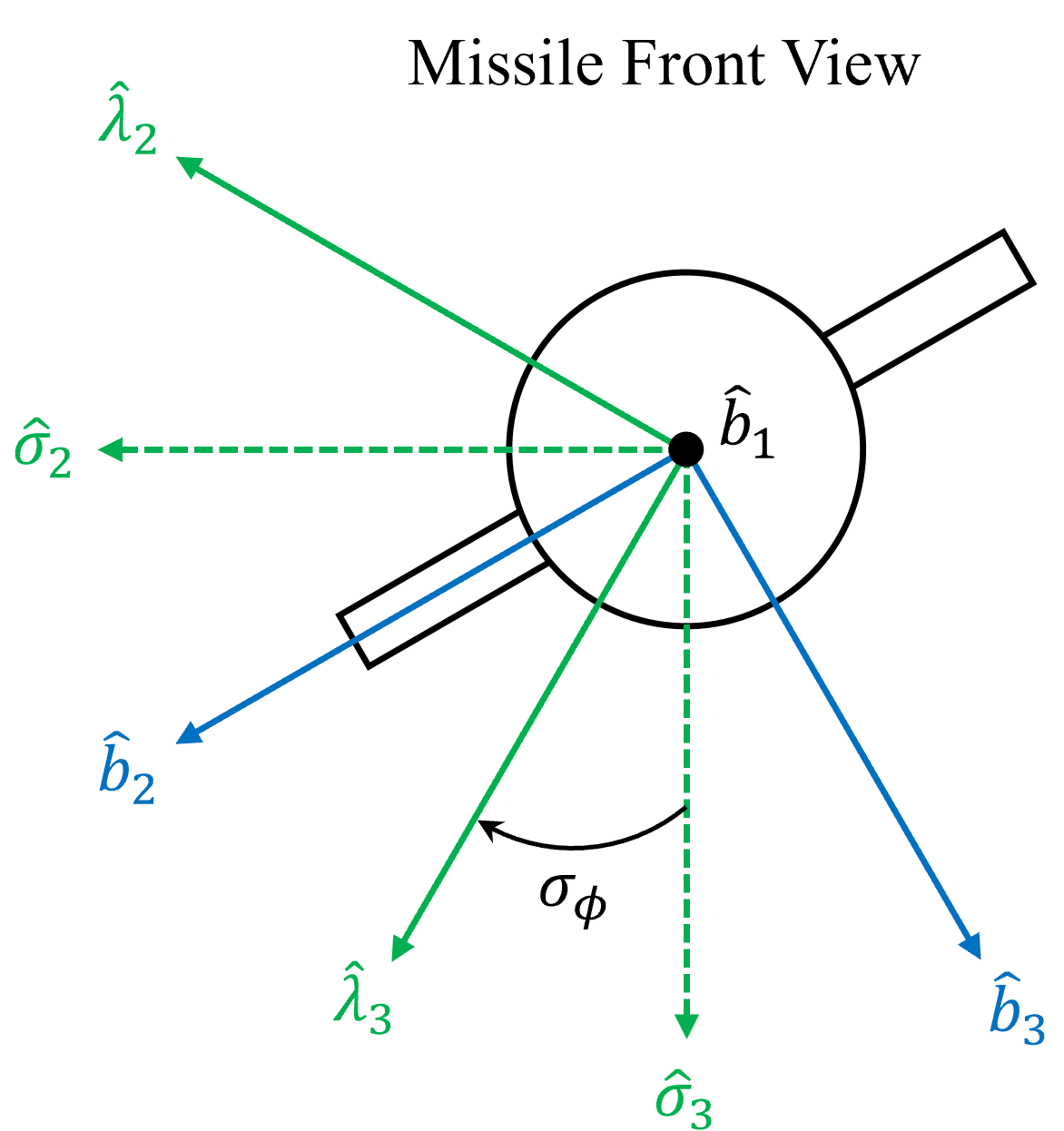}
  \caption{Definition of roll look angle}
  \label{fig:2_4_roll look angle}
\end{figure}

The roll look angle compensates for missile roll about $\hat{b}_1$-axis. With the target confined to the narrow seeker FOV, the \textit{\textbf{Assumption~\ref{assump:aoa}}} applies, allowing simplifications of both the roll look angle and the DCM between ${\sigma}$ and ${\lambda}$, as in (\ref{eq:2_roll_look_angle_roll})–(\ref{eq:2_DCM_look_los}). The roll look angle equals the roll (bank) angle in magnitude but opposite in sign, and its concept is illustrated in Fig. \ref{fig:2_4_roll look angle}. It can also be expressed using flight path, pitch, and yaw look angles by exploiting the consecutive transformation property of DCMs as in (\ref{eq:2_DCM_property}) \cite{c08}.

{\footnotesize
\begin{equation} \label{eq:2_DCM_4frames}
    C^\lambda_\sigma = C^\lambda_n C^n_b C^b_\sigma, \quad \text{or} \quad C^n_\lambda C^\lambda_\sigma = C^n_b C^b_\sigma
\end{equation}
}

By substituting equations (\ref{eq:2_DCM_nb}), (\ref{eq:2_dcm_nlambda}), and (\ref{eq:2_DCM_b_look}) into (\ref{eq:2_DCM_4frames}),

From the matrix elements $c_{32}$ and $c_{33}$ of (\ref{eq:2_DCM_4frames}), we have

{\footnotesize
\begin{equation}
    \begin{aligned}
    \sigma_\phi &= -\operatorname{atan2} \left( \frac{c_{32}}{c_{33}} \right) \\
    c_{32} &= s_\gamma s_{\sigma_\psi} + c_\gamma s_\mu c_{\sigma_\psi} \\
    c_{33} &= -s_\gamma c_{\sigma_\psi} s_{\sigma_\theta} 
    + c_\gamma s_\mu s_{\sigma_\psi} s_{\sigma_\theta} 
    + c_\gamma c_\mu c_{\sigma_\theta}
    \end{aligned}
\end{equation}
}

Based on the aforementioned small-angle assumption due to the narrow FOV, the LOS angles can be simplified as follows.

{\footnotesize
\begin{equation}
\label{eq:small_angle_los}
\begin{aligned}
    \lambda_\theta &= \theta + \sigma_\theta \approx \gamma + \sigma_\theta, \quad
\lambda_\psi = \psi + \sigma_\psi \approx \chi + \sigma_\psi
\end{aligned}
\end{equation} 
}

In (\ref{eq:small_angle_los}), $\theta$ and $\psi$ denote the missile body’s pitch and yaw angles with respect to the frame $\{n\}$, approximated by the flight path and heading angles under \textit{\textbf{Assumption~\ref{assump:aoa}}}. The defined angles, derivatives, DCMs, and angular velocities constitute the elements required for MPCG design, presented in Section~\ref{sec:2_guidance_algorithm}–\ref{subsec:MPCG}.

\subsection{Optimal Control Problem}

\begin{figure}[!htbp]
  \centering
  \includegraphics[width=0.9\columnwidth]{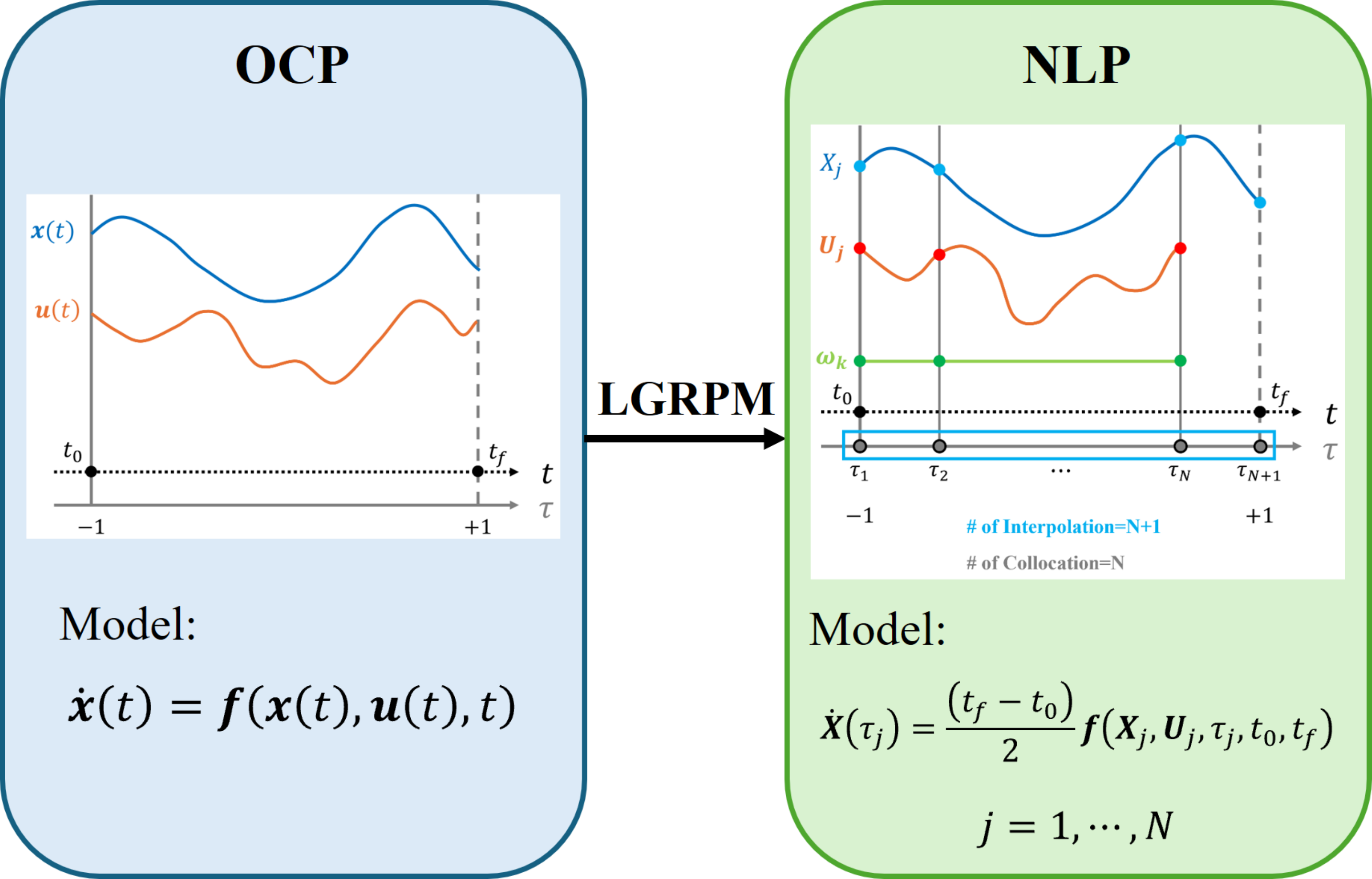}
  \caption{Transformation of the continuous-time OCP into a NLP problem via the LGRPM}
  \label{fig:2_7_LGRPM}
\end{figure}

The OCP seeks a control input $\bm{u}(t)$ that minimizes a performance index $J$ subject to system dynamics and constraints, typically expressed in Bolza form as

{\footnotesize
\begin{align}
    \min_{\bm{u}} \, J(\bm{u}) 
&= \phi\left( \bm{x}(t_0), \bm{x}(t_f), t_0, t_f \right) 
+ \int_{t_0}^{t_f} g\left( \bm{x}(t), \bm{u}(t), t \right) dt \nonumber \\
\text{\textbf{s.t.}} \quad
\dot{\bm{x}}(t) &= f \left( \bm{x}(t), \bm{u}(t), t \right) \label{eq:ocp_bolza} \\
\bm{c}_{\min} &\leq \bm{C}\left( \bm{x}(t), \bm{u}(t), t \right) \leq \bm{c}_{\max} \nonumber \\
\bm{\psi}_{\min} &\leq \bm{\psi} \left( \bm{x}(t_0), \bm{x}(t_f), t_0, t_f \right) \leq \bm{\psi}_{\max}  \nonumber
\end{align}
}

Since solving OCPs analytically is generally intractable, numerical methods are employed. Among them, direct collocation methods are widely adopted due to their computational efficiency and accuracy \cite{c14}.

In this study, the LGRPM is adopted to transcribe the continuous-time OCP into a finite-dimensional NLP problem. The LGRPM enforces system dynamics at collocation points and approximates states and controls by polynomial interpolation, providing accurate and efficient discretization of OCPs.

An additional advantage of LGRPM lies in the treatment of the terminal point: unlike other Gaussian collocation schemes, the differential equations are not enforced at the final interpolation node ($\tau=+1$), as illustrated in Fig. \ref{fig:2_7_LGRPM}. Instead, this point is used only for state interpolation. This characteristic is particularly suitable for MPC, since in a receding horizon framework the terminal state is not fixed a priori but updated dynamically at each step \cite{c15}.

Based on this structure, the LGRPM transforms the OCP into an NLP by 
following the steps below \cite{c26}:

{\footnotesize
\vspace{0.5ex}
\noindent \rule{\linewidth}{1.0pt}
\noindent \textbf{Algorithm 1:} \\
LGRPM Algorithm \\ 
\noindent \rule{\columnwidth}{0.5pt}
\noindent \textbf{1. Time Interval Normalization:}
\begin{align*}
 t = \frac{t_f - t_0}{2} \tau + \frac{t_f + t_0}{2}
\end{align*}

The time interval of the OCP, $t \in \left[ t_0, t_f \right]$, must be linearly transformed into a normalized interval, $\tau \in  \left[-1, 1 \right]$.

\vspace{2.0ex}
\noindent \textbf{2. LGR nodes Generation:} 
\begin{align*}
    \tau_1 < \tau_2 < \cdots < \tau_N < 1
\end{align*}

Using the Legendre polynomials, the N roots of $P_N(\tau) + P_{N-1}(\tau)$ are computed. These roots form the LGR points, which do not include the endpoint $(\tau_{N+1}=1)$. 

\vspace{2.0ex}
\noindent \textbf{3. Interpolation of States and Control Inputs:}
\begin{align*}
    \bm{x}(\tau) \approx \bm{X}(\tau) = \sum^{N+1}_{i=1} \bm{X}_i L^{(N)}_i (\tau) \\
    \bm{u}(\tau) \approx \bm{U}(\tau) = \sum^{N}_{j=1} \bm{U}_j L^{(N-1)}_j (\tau)
\end{align*}

The state variables are approximated at $N+1$ interpolation points $\tau_i$ using an $N$-th order Lagrange basis polynomial $L^{(N)}_i (\tau)$, where $\bm{X}_i=\bm{X} (\tau_i) \in \mathbb{R}^n$. Meanwhile, the control inputs are discretized at $N$ collocation points $\tau_j$ using $(N-1)$-th order Lagrange basis polynomial $L^{(N-1)}_{j} (\tau)$, where $\bm{U}_j=\bm{U} (\tau_j) \in \mathbb{R}^p$.

\vspace{2.0ex}
\noindent \textbf{4. Dynamic Constraint at LGR Collocation Points:}
\begin{align*}
    \frac{d \bm{x} (\tau)}{d\tau} &= \frac{(t_f - t_0)}{2} \bm{f} (\bm{x}(\tau), \bm{u}(\tau), \tau, t_0, t_f) \\
    \Rightarrow \dot{\bm{X}}(\tau_j) &\approx \sum^{N+1}_{i=1} \bm{X}_i \dot{L}^{(N)}_{i} (\tau_j) \\
    &= \sum^{N+1}_{i=1} D_{ji} \bm{X}_i (\tau_j) \\
    &= \frac{(t_f - t_0)}{2} \bm{f} (\bm{X}_j, \bm{U}_j, \tau_j, t_0, t_f), 
    \quad j = 1, \cdots, N
\end{align*}

At the $N$ LGR collocation points, the approximated state variables must exactly satisfy the equations of motion. To enforce the system dynamics as constraints, the derivatives of Lagrange basis polynomial $\dot{L}^{(N)}_{i}$ and the differentiation matirx $\left[ D_{ji} \right]$ are applied, where $\left[ D_{ji} \right] \in \mathbb{R}^{N \times (N+1)}$. 

\vspace{2.0ex}
\noindent \textbf{5. Approximation of Objective Function:}
\begin{align*}
    &\min_u J = \\
    &\phi(\bm{x}(-1), \bm{x}(1), t_0, t_f) + \frac{(t_f - t_0)}{2} \int^{1}_{-1} g(\bm{x}(\tau), \bm{u}(\tau), \tau, t_0, t_f) d\tau \\[1ex]
    \Rightarrow &\min_{\bm{u}_{\tau_1:\tau_N}} J = \\
    &\phi(\bm{X}_1, \bm{X}_{N+1}, t_0, t_f) + \frac{(t_f - t_0)}{2} \sum^N_{j=1} \omega_j g(\bm{X}_j, \bm{U}_j, \tau_j, t_0, t_f)
\end{align*}

The objective function, which involves integration, is approximated using Gaussian quadrature based on the $N$ LGR collocation points. Here, $\omega_j$ is the corresponding quadrature weights.

\vspace{2.0ex}
\noindent \textbf{6. Path Constraints and Boundary Conditions:}
\begin{align*}
    \frac{(t_f - t_0)}{2}\bm{C}(\bm{x}(\tau), \bm{u}(\tau), \tau, t_0, t_f) \leq 0 \\
    \psi(\bm{x}(-1), \bm{x}(1), t_0, t_f) = 0
\end{align*}

\begin{align*}
    \Rightarrow \frac{(t_f - t_0)}{2}\bm{C}(\bm{X}_j, \bm{U}_j, \tau_j, t_0, t_f) \leq 0 \\
    \psi(\bm{X}_1, \bm{X}_{N+1}, t_0, t_f) = 0 \\
    where, \quad j = 1, \cdots, N
\end{align*}

The path constraints are enforced at the $N$ LGR collocation points, while the boundary conditions are satisfied at the initial and final time points.

\vspace{2.0ex}
\noindent \textbf{7. NLP Problem Setup and Optimization:}
\begin{align*}
    &\min_{\bm{u}_{\tau_1:\tau_N}} J = \\ &\phi(\bm{X}_1, \bm{X}_{N+1}, t_0, t_f) + \frac{(t_f - t_0)}{2} \sum^N_{j=1} \omega_j g(\bm{X}_j, \bm{U}_j, \tau_j, t_0, t_f)
\end{align*}

\textbf{s.t.}

\begin{align*}
    \sum^{N+1}_{i=1} D_{ji} \bm{X}_i (\tau_j)
    = \frac{(t_f - t_0)}{2} \bm{f} &(\bm{X}_j, \bm{U}_j, \tau_j, t_0, t_f) \\
    \frac{(t_f - t_0)}{2}\bm{C}(\bm{X}_j, \bm{U}_j, \tau_j, t_0, t_f) &\leq 0 \\
    \psi(\bm{X}_1, \bm{X}_{N+1}, t_0, t_f) &= 0, \\
    where, \quad j = 1, \cdots, N
\end{align*}

\noindent \rule{\columnwidth}{0.5pt}
}

Based on the discretized objective function, dynamic constraints, path constraints, and boundary conditions, the OCP is reformulated as an NLP. To solve this NLP, several optimization solvers are commonly used, including the Interior Point Optimized (IPOPT) \cite{c27}, Sequential Quadratic Programming (SQP) methods, and the Sparse Nonlinear Optimizer (SNOPT).

\subsection{A Look Angle-Based Nonlinear Model Predictive Control for Guidance Law} \label{subsec:MPCG}

MPC is a control strategy that predicts future system behavior and computes an optimal sequence of control inputs over a finite horizon, while only the first input is applied at each step. The process is repeated as the system state is updated, which defines the receding horizon strategy \cite{c19, c20}.

\begin{figure}[!htbp]
  \centering
  \includegraphics[width=0.9\columnwidth]{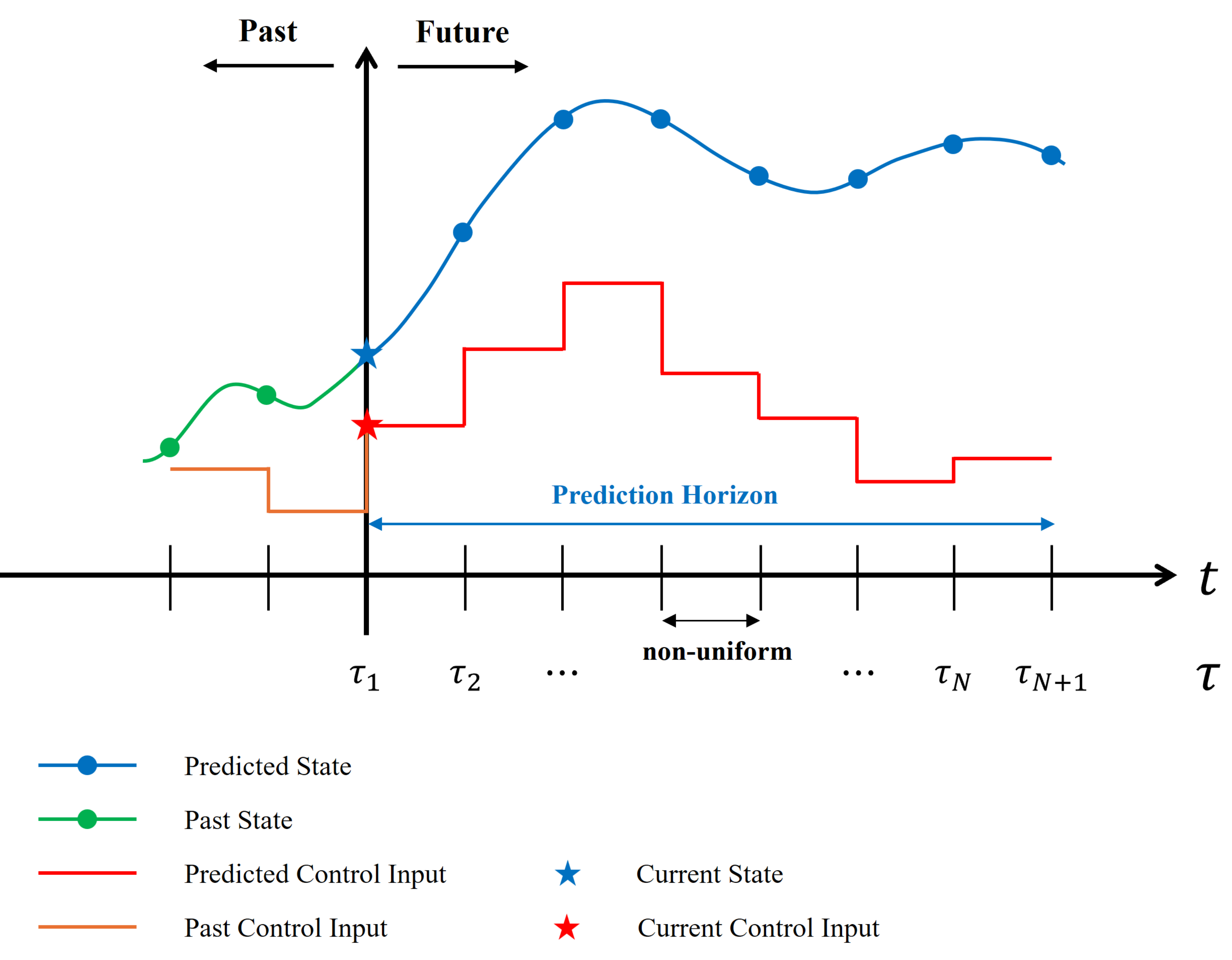}
  \caption{Operational concept of MPC using LGRPM}
  \label{fig:2_8_MPC}
\end{figure}

As shown in Fig. \ref{fig:2_8_MPC}, the current state (blue star) is used with the system’s dynamic model to predict future trajectories discretized by non-uniform interpolation points $\tau_1, \dots, \tau_{N+1}$. An optimal input sequence is then obtained, but only the input at the first collocation point (red star) is applied.

MPC enhances system performance in dynamic environments by continually updating control inputs and systematically incorporating constraints, offering distinct advantages for guidance law design:

\begin{enumerate}
    \item MPC can maintain effective performance even under highly dynamic engagement scenarios between a missile and a target.

    \item The seeker’s FOV limitations — such as allowable look angle $\sigma_{FOV}$ and detection range $R_{FOV}$  — can be explicitly considered during the controller design process.

    \item Physical limitations, such as the missile's maximum allowable acceleration, can be imposed as constraints to ensure that the control inputs remain within feasible operational boundaries.
\end{enumerate}

The FOV of a missile seeker, characterized by the maximum allowable look angle $\sigma_{FOV}$ and the detection range $R_{FOV}$, as shown in Fig. \ref{fig:1_1_comparison_seekers}-(b), imposes significant constraints on guidance performance. MPCG provides enhanced flexibility by explicitly incorporating these constraints into the guidance law, thereby offering advantages over conventional methods such as PPNG.

\begin{align}
\bm{x} &= 
\left[
\bm{r}^b,\, \bm{v}^b,\, \sigma_\theta,\, \sigma_\psi,\, \dot{\sigma}_\theta,\, \dot{\sigma}_\psi,\, \tilde{\Omega}_\theta^b,\, \tilde{\Omega}_\psi^b
\right]^{\top} \label{eq:2_mpcg_state} \\
\bm{u} &= 
\left[\bm{a}_M^b \right]^{\top} \nonumber \\
\text{where,} \quad 
&\bm{x} \in \mathbb{R}^{n \times 1}, \quad 
\bm{u} \in \mathbb{R}^{p \times 1} \nonumber \\
\min_{\bm{u}} J &= \bm{x}_{N+1}^{\top} \bm{Q}_f \bm{x}_{N+1} 
+ \sum_{j=1}^{N} \left( \bm{x}_j^{\top} \bm{Q} \bm{x}_j + \bm{u}_j^{\top} \bm{R} \bm{u}_j \right) \label{eq:2_mpcg_object} \\
\text{here,} \quad 
&\bm{Q},\, \bm{Q}_f \in \mathbb{R}^{n \times n}, \quad 
\bm{R} \in \mathbb{R}^{p \times p} \nonumber \\
\text{\textbf{s.t.}} \quad \nonumber \\
&\begin{aligned}[t] \label{eq:2_mpcg_const}
    \dot{\bm{x}} &= f(\bm{x}, \bm{u}) \\
    \left\| \sigma_m \right\| &\leq \sigma_{FOV} \\
    \left\| \dot{\sigma}_{\theta,\psi} \right\| &\leq \dot{\sigma}_{FOV} \\
    \left\| \bm{a}_M^b \right\| &\leq a_{\max} \\
    R &\leq R_{FOV} \\
    a_{1,m}^b &= 0
\end{aligned}
\end{align}

In addition, when integrated with a conventional three-loop autopilot, the use of acceleration constraints in MPCG may help alleviate potential integrator windup issues, which typically arise under actuator saturation and lead to persistent error accumulation. By ensuring that control commands remain within predefined bounds, MPCG reduces the likelihood of such problems and provides a structurally robust framework for guidance under constraints. Although direct verification of windup mitigation is beyond the scope of this study, it is reasonable to expect that the inclusion of acceleration constraints can contribute to alleviating its effects  \cite{c28}.

\begin{figure}[!htbp]
  \centering
  \includegraphics[width=0.95\linewidth]{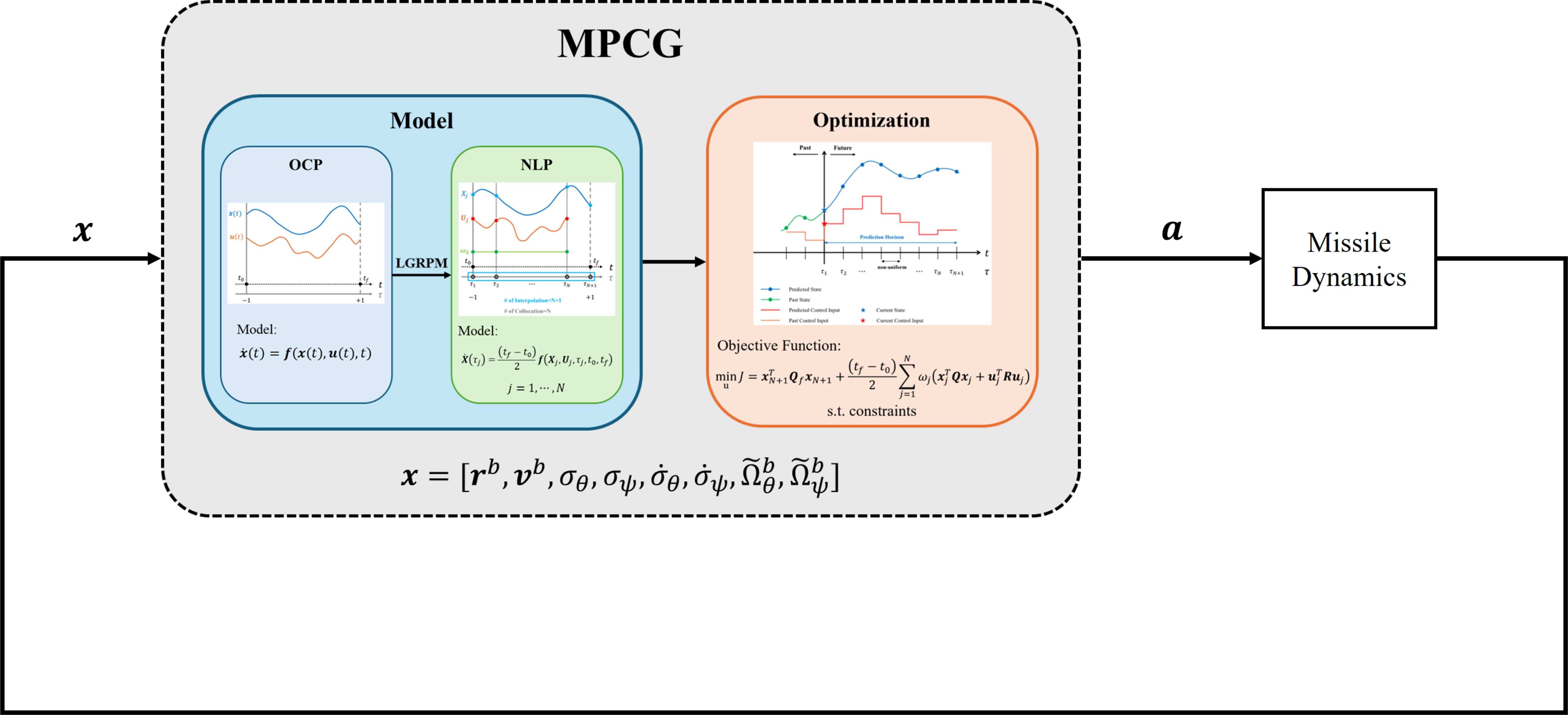}
  \caption{Simplified framework of MPCG}
  \label{fig:2_9_simple_MPCG}
\end{figure}

To facilitate understanding of how MPCG operates, Fig. \ref{fig:2_9_simple_MPCG} illustrates the simplified framework of the MPC-based guidance system.

At each control step, the state vector is measured or estimated, and the MPCG discretizes the continuous-time OCP into an NLP using LGRPM. The resulting formulation incorporates dynamic, path, and boundary constraints while minimizing a quadratic cost function. Following the receding horizon strategy, only the first control input of the optimized sequence is applied as the missile’s acceleration command.

Building upon this framework, the mathematical formulation of the MPC problem is given as follows:

Equations (\ref{eq:2_mpcg_state})–(\ref{eq:2_mpcg_const}) summarize the formulation of the MPCG problem. The state vector $\bm{x}$ consists of the relative position and velocity in the body-fixed frame, the pitch and yaw look angles with their rates, and the simplified LOS angular velocities $(\tilde{\Omega}^b_\theta, \tilde{\Omega}^b_\psi)$. The control input $\bm{u}$ is defined as the acceleration vector of the missile in the body-fixed frame.

The optimization problem is formulated in the quadratic form shown in (\ref{eq:2_mpcg_object}), where the weighting matrices $\bm{Q}$, $\bm{Q}_f$, and $\bm{R}$ regulate the contributions of the states and control inputs. The dynamics and constraints are expressed in (\ref{eq:2_mpcg_const}), which enforce the LOS angle limits, look angle rate bounds, acceleration saturation, and seeker detection range. Additionally, the longitudinal acceleration along the body x-axis is set to zero, consistent with the assumptions of the missile model.

\begin{align} 
\min_{\bm{u}_{\tau_1:\tau_N}} J
&= \bm{x}_{N+1}^{\top}  \bm{Q}_f \bm{x}_{N+1} \\ &\qquad + \frac{t_f - t_0}{2} \sum_{j=1}^{N} \omega_j  
\left( \bm{x}_j^{\top} \bm{Q} \bm{x}_j + \bm{u}_j^{\top} \bm{R} \bm{u}_j \right)  \nonumber \\
\text{\textbf{s.t.}} \quad \nonumber \\
&\begin{aligned} \label{eq:2_mpcg_const_2}
    \dot{\bm{r}}^b &\approx \bm{v}^b \\
    \dot{\bm{v}}^b &\approx \bm{a}^b = \bm{a}_T^b - \bm{a}_M^b \\
    \frac{d \sigma_\theta}{dt} &= \dot{\sigma}_\theta, \quad
    \frac{d \sigma_\psi}{dt} = \dot{\sigma}_\psi \\
    \frac{d \dot{\sigma}_\theta}{dt} &= \ddot{\sigma}_\theta, \quad
    \frac{d \dot{\sigma}_\psi}{dt} = \ddot{\sigma}_\psi \\
    \frac{d \tilde{\Omega}_\theta^b}{dt} &= \dot{\tilde{\Omega}}_\theta^b, \quad
    \frac{d \tilde{\Omega}_\psi^b}{dt} = \dot{\tilde{\Omega}}_\psi^b
\end{aligned} \\
&\begin{aligned} \label{eq:2_mpcg_const_3}
    \cos{\sigma_\theta} \cos{\sigma_\psi} &\geq \cos{\sigma_{FOV}} \\
    \dot{\sigma}_{\theta,\psi}^2 &\leq \dot{\sigma}_{FOV}^2 \\
    R^2 &\leq R_{FOV}^2 \\
    (\bm{a}_M^b)^{\top} \cdot \bm{a}_M^b &\leq a_{\max}^2 \\
    a_{1,m}^b(\tau_j) &= 0, \hspace{3.35em} j = 1, \dots, N \\
    \bm{a}_T^b(\tau_j) &= \bm{a}_T^b(\tau_1), \quad j = 2, \dots, N
\end{aligned}
\end{align}

The objective function of the LGRPM-based MPC formulation is defined in (\ref{eq:2_mpcg_const_2}), where the weighting matrices $\bm{Q}, \bm{Q}_f, \bm{R}$ regulate the performance and sensitivity of the control, and $\omega_j$ denotes the quadrature weights associated with the LGR collocation points.

The dynamic constraints in (\ref{eq:2_mpcg_const_2}) establish the kinematic relationships among the relative position, velocity, and acceleration in the body-fixed frame, while also specifying the time derivatives of the look angles, their rates, and the simplified LOS angular velocities.

Finally, the path constraints are imposed through (\ref{eq:2_mpcg_const_3}), ensuring that the total look angle defined in (\ref{eq:2_total_look_angle}), the relative range, and the missile acceleration remain within admissible bounds. To reduce computational burden, these constraints are expressed in squared form. Moreover, the x-component of the missile’s acceleration is fixed to zero at all collocation points, while the target acceleration is uniformly applied across the prediction horizon.

Specifically, The time derivatives of the pitch and yaw look angle rates $(\ddot{\sigma}_{\theta, \psi})$ in (\ref{eq:2_mpcg_const_2}) are derived from the definition of the look angle rates given in (\ref{eq:2_look_angle_rates}). The resulting expression is presented below.

\begin{align} \label{eq:2_rb_vb_def}
\bm{r}^b &= r_1^b \hat{b}_1 + r_2^b \hat{b}_2 + r_3^b \hat{b}_3 
= \begin{bmatrix} r_1^b \\ r_2^b \\ r_3^b \end{bmatrix}
\triangleq \begin{bmatrix} r_1 \\ r_2 \\ r_3 \end{bmatrix} \nonumber \\
\bm{v}^b &= v_1^b \hat{b}_1 + v_2^b \hat{b}_2 + v_3^b \hat{b}_3 
= \begin{bmatrix} v_1^b \\ v_2^b \\ v_3^b \end{bmatrix}
\triangleq \begin{bmatrix} \dot{r}_1 \\ \dot{r}_2 \\ \dot{r}_3 \end{bmatrix} \\
\bm{a}^b &= a_1^b \hat{b}_1 + a_2^b \hat{b}_2 + a_3^b \hat{b}_3 
= \begin{bmatrix} a_1^b \\ a_2^b \\ a_3^b \end{bmatrix}
\triangleq \begin{bmatrix} \ddot{r}_1 \\ \ddot{r}_2 \\ \ddot{r}_3 \end{bmatrix} \nonumber
\end{align}

{\footnotesize
\begin{align} \label{eq:2_R_R2}
R &= \sqrt{r_1^2 + r_2^2 + r_3^2}, \quad
R_\perp = \sqrt{r_1^2 + r_2^2} \nonumber \\
e_1 &= r_1 \dot{r}_1 + r_2 \dot{r}_2, \qquad
e_2 = r_1 \ddot{r}_1 + r_2 \ddot{r}_2 \\
e_3 &= r_1 \dot{r}_1 + r_2 \dot{r}_2 + r_3 \dot{r}_3 \nonumber \\
e_4 &= r_1 \dot{r}_2 - r_2 \dot{r}_1, \qquad
e_5 = r_1 \ddot{r}_2 - r_2 \ddot{r}_1 \nonumber \\[1ex]
E_1 &= r_3 ( \dot{r}_1^2 + \dot{r}_2^2 + e_2 ) - \dot{r}_3 e_1 - \ddot{r}_3 R_\perp^2 \nonumber \\
E_2 &= R^2 R_\perp, \qquad
E_3 = r_3 e_1 - \dot{r}_3 R_\perp^2 \\
E_4 &= 2 e_3 R_\perp + e_1 \frac{R^2}{R_\perp}, \qquad
E_5 = e_5 R_\perp^2 - 2 e_1 e_4 \nonumber
\end{align}
}

\begin{equation} \label{eq:2_ddot_look_angles}
\begin{aligned}
\ddot{\sigma}_\theta &= \frac{E_1 E_2 - E_3 E_4}{(R^2 R_\perp)^2}, \quad
\ddot{\sigma}_\psi = \frac{E_5}{(R_\perp^2)^2}
\end{aligned}
\end{equation}

Subsequently, the time derivatives of the simplified pitch and yaw LOS angular velocities $\dot{\tilde{\Omega}}^b_{\theta, \psi}$ expressed in the body-fixed frame, as introduced in (\ref{eq:2_mpcg_const_2}), are derived from the LOS angular velocity defined in the LOS frame in (\ref{eq:2_LOS_angular_vel}).

By applying the DCM $C^b_\lambda$ in (\ref{eq:2_LOS_angular_vel}), the LOS angular velocity in the body-fixed frame can be expressed as shown in (\ref{eq:2_LOS_angular_vel_body}), where the pitch and yaw components are formulated in terms of the DCM and the LOS rates $\dot{\lambda}_\theta, \dot{\lambda}_\psi$.

\begin{equation} \label{eq:2_LOS_angular_vel_body}
    \bm{\Omega}^b = C^b_\lambda \bm{\Omega}^\lambda = 
    C^b_\lambda \begin{bmatrix}
        0 \\ \dot{\lambda}_\theta \\ \dot{\lambda}_\psi c_{\lambda_\theta}
    \end{bmatrix}
    = \begin{bmatrix}
        \Omega^b_\phi \\[0.3ex] \Omega^b_\theta \\[0.3ex] \Omega^b_\psi
    \end{bmatrix}
\end{equation}

By substituting (\ref{eq:2_DCM_b_look}) and (\ref{eq:2_DCM_look_los}) into (\ref{eq:2_LOS_angular_vel_body}),

\begin{align}
\begin{bmatrix}
\Omega^b_\phi \\
\Omega^b_\theta \\
\Omega^b_\psi
\end{bmatrix}
&= C^b_\lambda
\begin{bmatrix}
0 \\
\dot{\lambda}_\theta \\
\dot{\lambda}_\psi c_{\lambda_\theta} 
\end{bmatrix}
= C^b_\sigma C^\sigma_\lambda
\begin{bmatrix}
0 \\
\dot{\lambda}_\theta \\
\dot{\lambda}_\psi c_{\lambda_\theta} 
\end{bmatrix} \label{eq:2_los_angular_vel_induce} \\
&=
\begin{bmatrix}
\Omega^b_\phi \\
\Omega^b_{\theta 1}
+ \Omega^b_{\theta 2}  \\
\dot{\lambda}_\theta c_{\sigma_\theta} s_{\sigma_\phi} 
+ \dot{\lambda}_\psi c_{\lambda_\theta} c_{\sigma_\theta} c_{\sigma_\phi}
\end{bmatrix} \nonumber\
\end{align}

\begin{equation} \label{eq:2_los_angular_vel_body}
\begin{aligned}
\Omega^b_\theta &= 
\Omega^b_{\theta 1} + \Omega^b_{\theta 2}, \quad
\Omega^b_\psi = 
\dot{\lambda}_\theta c_{\sigma_\theta} s_{\sigma_\phi}
+ \dot{\lambda}_\psi c_{\lambda_\theta} c_{\sigma_\theta} c_{\sigma_\phi} \\[2ex]
\Omega^b_{\theta 1} &= \dot{\lambda}_\theta 
\left( 
c_{\sigma_\psi} c_{\sigma_\phi} 
+ s_{\sigma_\psi} s_{\sigma_\theta} s_{\sigma_\phi} 
\right) \\
\Omega^b_{\theta 2} &= \dot{\lambda}_\psi c_{\lambda_\theta} 
\left( 
s_{\sigma_\psi} s_{\sigma_\theta} c_{\sigma_\phi}
- c_{\sigma_\psi} s_{\sigma_\phi}
\right)
\end{aligned}
\end{equation}

It follows from (\ref{eq:2_los_angular_vel_body}) that explicit expressions for the pitch and yaw components in the body-fixed frame can be obtained.

However, since the LOS rates are not included in the state vector defined in (\ref{eq:2_mpcg_state}), the angular velocities $\Omega^b_\theta$ and $\Omega^b_\psi$ must be reformulated in terms of the look angles and their time derivatives. To achieve this consistency within the control framework, the angular velocity composition rule given in (\ref{eq:2_los_composite}) is applied.

\begin{equation} \label{eq:2_los_composite}
    \bm{\omega}^b_{n \lambda} = \bm{\omega}^b_{n b} + \bm{\omega}^b_{b \sigma} + \bm{\omega}^b_{\sigma \lambda}
\end{equation}

Using the DCM, this expression can be represented as follows.

\begin{equation} \label{eq:2_sum_angular_vel}
    C^b_\sigma C^\sigma_\lambda \bm{\omega}^\lambda_{n \lambda} = \bm{\omega}^b_{n b} + \bm{\omega}^b_{b \sigma} + C^b_\sigma \bm{\omega}^\sigma_{\sigma \lambda}
\end{equation}

By substituting (\ref{eq:2_omega_b_nb}), (\ref{eq:2_angular_vel_n_los2}), (\ref{eq:2_DCM_b_look}), (\ref{eq:2_angular_vel_body_look}), (\ref{eq:2_DCM_look_los}), and (\ref{eq:2_angular_vel_look_los}) into (\ref{eq:2_sum_angular_vel}), the LOS rates are expressed in terms of the look angles and their rates, as summarized in (\ref{eq:2_LOS_rates_2}).

\begin{figure*}[!htbp]
  \centering
  \includegraphics[width=\textwidth]{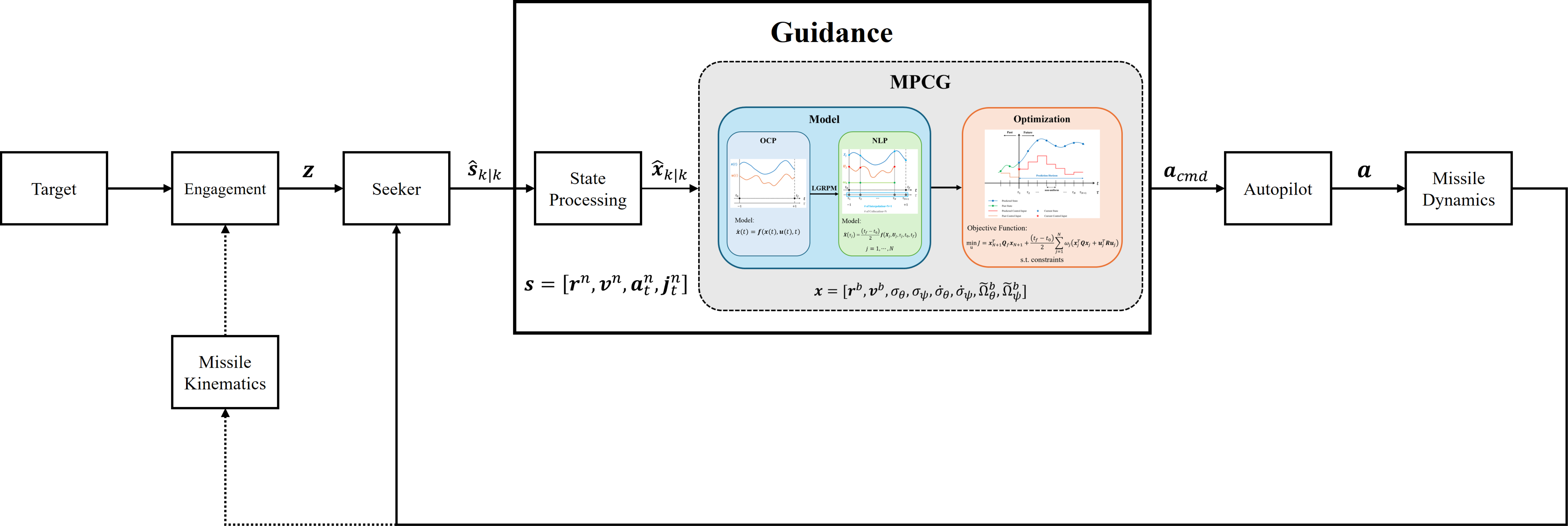}
  \caption{Schematic of MPCG-based missile GNC loop}
  \label{fig:2_10_framework_mpcg}
\end{figure*}

{\footnotesize
\begin{equation} \label{eq:2_LOS_rates_2}
\begin{aligned}
\dot{\lambda}_\theta &= 
\frac{
\dot{\gamma} + \dot{\sigma}_\theta c_{\sigma_\psi} + \dot{\sigma}_\phi s_{\sigma_\psi} c_{\sigma_\theta}
}{
c_{\sigma_\psi} c_{\sigma_\phi} + s_{\sigma_\psi} s_{\sigma_\theta} s_{\sigma_\phi}
} \\
\dot{\lambda}_\psi &= 
\frac{
\dot{\chi} c_{\gamma} + \dot{\sigma}_\psi 
- \dot{\sigma}_\phi s_{\sigma_\theta} 
- \dot{\lambda}_\theta c_{\sigma_\theta} s_{\sigma_\phi}
}{
c_{\gamma}
}
\end{aligned}
\end{equation}
}

To simplify (\ref{eq:2_los_angular_vel_body}) and (\ref{eq:2_LOS_rates_2}), the following assumptions are introduced.






\begin{assumption}\label{assump:fov} 
The strapdown seeker has a narrow FOV, and the MPCG ensures that the look angles remain within this range, allowing the small-angle assumption to be applied.  
\end{assumption}

\vspace{0.2ex}

\begin{assumption}\label{assump:stt} 
For a Skid-To-Turn (STT) missile, the roll angle is controlled to stay near zero, so the roll look angle and its rate are assumed negligible in the guidance formulation.  
\end{assumption}

Based on \textbf{\textit{Assumption \ref{assump:fov}}}  and \textbf{\textit{\ref{assump:stt}}}, (\ref{eq:2_los_angular_vel_body}) and (\ref{eq:2_LOS_rates_2}) can be rewritten as follows.

{\footnotesize
\begin{equation} \label{eq:2_LOS_angular_vel_b_approx}
\begin{aligned}
    \Omega^b_\theta
    &= \dot{\lambda}_\theta (c_{\sigma_\psi} c_{\sigma_\phi} + s_{\sigma_\psi} s_{\sigma_\theta} \bm{s_{\sigma_\phi}}) 
    + \dot{\lambda}_\psi c_{\lambda_\theta}(s_{\sigma_\psi} s_{\sigma_\theta} c_{\sigma_\phi} - c_{\sigma_\psi} \bm{s_{\sigma_\phi}}) \\
    &\approx \dot{\lambda}_\theta c_{\sigma_\psi} + \dot{\lambda}_\psi c_{\lambda_\theta} s_{\sigma_\psi} s_{\sigma_\theta} \\[1ex]
    \Omega^b_\psi 
    &= \dot{\lambda}_\theta c_{\sigma_\theta} \bm{s_{\sigma_\phi}} + \dot{\lambda}_\psi c_{\lambda_\theta} c_{\sigma_\theta} c_{\sigma_\phi} 
    \approx \dot{\lambda}_\psi c_{\lambda_\theta} c_{\sigma_\theta}
\end{aligned}   
\end{equation}
}

{\footnotesize
\begin{equation} \label{eq:2_look_angle_rates_approx}
\begin{aligned}
\dot{\lambda}_\theta 
&= \frac{\dot{\gamma} + \dot{\sigma}_\theta c_{\sigma_\psi} + \dot{\sigma}_\phi s_{\sigma_\psi} c_{\sigma_\theta}}{c_{\sigma_\psi} c_{\sigma_\phi} + s_{\sigma_\psi} s_{\sigma_\theta} s_{\sigma_\phi}} 
\approx \frac{\dot{\gamma} + \dot{\sigma}_\theta c_{\sigma_\psi}}{c_{\sigma_\psi}} \\
\dot{\lambda}_\psi
&= \frac{\dot{\chi} c_{\gamma} + \dot{\sigma}_\psi - \dot{\sigma}_\phi s_{\sigma_\theta} - \dot{\lambda}_\theta c_{\sigma_\theta} s_{\sigma_\phi}}{c_{\gamma}} 
\approx \frac{\dot{\chi} c_{\gamma} + \dot{\sigma}_\psi}{c_{\gamma}} 
\end{aligned}
\end{equation}
}

By applying (\ref{eq:small_angle_los}) to $c_{\lambda_\theta}$ in (\ref{eq:2_LOS_angular_vel_b_approx}), along with the trigonometric identities and \textbf{\textit{Assumption \ref{assump:fov}}}, the expression can be rewritten as follows.

{\footnotesize
\begin{equation} \label{eq:2_cos_los}
\begin{aligned}
   \cos{\lambda_\theta} &\approx \cos(\gamma + \sigma_\theta) \\
&= \cos \gamma \cos \sigma_\theta - \sin \gamma \sin \sigma_\theta  \\
&\approx \cos \gamma \cos \sigma_\theta  
\end{aligned}
\end{equation}
}

By substituting (\ref{eq:2_look_angle_rates_approx}) and (\ref{eq:2_cos_los}) into (\ref{eq:2_LOS_angular_vel_b_approx}),

{\footnotesize
\begin{equation} \label{eq:2_LOS_angular_vel_b_approx_2}
\begin{aligned} 
\Omega^b_\theta 
&= \dot{\lambda}_\theta c_{\sigma_\psi} 
+ \dot{\lambda}_\psi c_{\lambda_\theta} s_{\sigma_\psi} s_{\sigma_\theta} \\
&\approx 
\left( \frac{ \dot{\gamma} + \dot{\sigma}_\theta c_{\sigma_\psi} }{ c_{\sigma_\psi} } \right) c_{\sigma_\psi} 
+ \left( \frac{ \dot{\chi} c_\gamma + \dot{\sigma}_\psi }{ c_\gamma } \right)
c_\gamma c_{\sigma_\theta} s_{\sigma_\psi} s_{\sigma_\theta}    \\
&= \dot{\gamma} + \dot{\sigma}_\theta c_{\sigma_\psi}
+ \left( \dot{\chi} c_\gamma + \dot{\sigma}_\psi \right)
c_{\sigma_\theta} s_{\sigma_\psi} s_{\sigma_\theta}  \\[2ex]
\Omega^b_\psi 
&= \dot{\lambda}_\psi c_{\lambda_\theta} c_{\sigma_\theta}
\approx \left( \frac{ \dot{\chi} c_\gamma + \dot{\sigma}_\psi }{ c_\gamma } \right)
c_\gamma c_{\sigma_\theta} c_{\sigma_\theta}  \\
&= \dot{\chi} c_\gamma c^2_{\sigma_\theta}
+ \dot{\sigma}_\psi c^2_{\sigma_\theta} 
\end{aligned}
\end{equation}
}

In (\ref{eq:2_LOS_angular_vel_b_approx_2}), the time derivatives of the flight path angle and heading angle are not included as control variables in the MPCG formulation. Therefore, these terms are disregarded to simplify the expressions.

\begin{equation}
\begin{aligned}
    \tilde{\Omega}^b_\theta &\triangleq \dot{\sigma}_\theta c_{\sigma_\psi} + \dot{\sigma}_\psi s_{\sigma_\psi} s_{\sigma_\theta} c_{\sigma_\theta} \\
    \tilde{\Omega}^b_\psi &\triangleq \dot{\sigma}_\psi c^2_{\sigma_\theta}
\end{aligned}
\end{equation}

Then, the time derivatives of the simplified pitch and yaw LOS angular velocities $\dot{\tilde{\Omega}}^b_{\theta, \psi}$ for dynamic constraints of the MPCG in (\ref{eq:2_mpcg_const_2}).

{\footnotesize
\begin{equation} \label{simplified_LOS_angular_rates}
\begin{aligned}
    \dot{\tilde{\Omega}}^b_\theta 
    &= \ddot{\sigma}_\theta c_{\sigma_\psi} 
    - \dot{\sigma}_\psi \dot{\sigma}_\theta s_{\sigma_\psi}
    + \ddot{\sigma}_\psi s_{\sigma_\psi} s_{\sigma_\theta} c_{\sigma_\theta} 
    + \dot{\sigma}_\psi^2 c_{\sigma_\psi} s_{\sigma_\theta} c_{\sigma_\theta} \\
    &\quad + \dot{\sigma}_\psi \dot{\sigma}_\theta s_{\sigma_\psi} c^2_{\sigma_\theta} 
    - \dot{\sigma}_\psi \dot{\sigma}_\theta s_{\sigma_\psi} s^2_{\sigma_\theta} \\[2ex]
    \dot{\tilde{\Omega}}^b_\psi 
    &= \ddot{\sigma}_\psi c^2_{\sigma_\theta} 
    - 2 \dot{\sigma}_\psi \dot{\sigma}_\theta c_{\sigma_\theta} s_{\sigma_\theta} 
\end{aligned}
\end{equation}
}

The time derivatives of the pitch and yaw look angle rates $\ddot{\sigma}_{\theta, \psi}$ in (\ref{simplified_LOS_angular_rates}) has already been defined in (\ref{eq:2_ddot_look_angles}). With this, the definitions of the state variables and the formulation of the dynamic constraints required for implementing MPCG have been completed.

\subsection{Summary}

Fig. \ref{fig:2_10_framework_mpcg} depicts the schematic of the MPCG-based missile GNC loop, illustrating the flow of information from seeker measurements to guidance commands within the MPC framework. The seeker provides observational data, which are processed into the state variables required for optimization. Two types of states are distinguished: the measured state vector $\hat{\bm{s}}_{k|k}$, obtained directly from the seeker and preprocessing, and the augmented state vector $\bm{x}$, formulated for the MPCG problem. While, $\bm{s}$ contains only the relative position, velocity, and target-related information in the navigation frame, $\bm{x}$ additionally incorporates look angles, their rates, and simplified LOS angular velocities expressed in the missile body-fixed frame, ensuring consistency with the MPC formulation.

The control problem is then posed as a quadratic optimization over the prediction horizon, subject to dynamic and path constraints. By appropriately selecting the weighting matrices, the controller can prioritize critical states such as relative position or acceleration while suppressing less influential terms. Of particular importance is the minimization of the look angles, their rates, and the simplified LOS angular velocities. Small values of these quantities ensure that the missile consistently points toward the target, avoids excessive oscillations, and maintains stable tracking. Moreover, reducing the LOS angular velocities implies that the missile’s body is aligned with the LOS vector, leading to a near-ideal intercept trajectory. This makes their minimization a key component of the effectiveness of the MPCG strategy.


\section{Target Input Estimation} \label{sec:3_target_estimation}

\subsection{Adaptive Extended Kalman Filter}

The AEKF is adopted to handle the noisy and uncertain measurements characteristic of missile–target engagements, thereby enhancing the stability and effectiveness of MPC-based guidance. By adaptively tuning the process and measurement noise covariances in online, the AEKF provides more accurate state estimation compared with the conventional EKF.

{\footnotesize
\begin{equation} \label{eq:3_kf_diff_eq}
\begin{aligned}
    x_{k+1} &= F_k x_k + G_k u_k + w_k \\
    z_k &= h_k (x_k) + v_k 
\end{aligned}
\end{equation} 
}

The state-space representation of the AEKF is given in (\ref{eq:3_kf_diff_eq}), where $x_k, u_k, z_k$ denote the state, input, and measurement vectors at time step $k$. Here, $F_k$ is the state transition matrix, $G_k$ the input matrix, and $h_k (\cdot)$ a nonlinear measurement function. The dimensions are defined as $x \in \mathbb{R}^m, u \in \mathbb{R}^r,$ and $ z \in \mathbb{R}^q$. The process noise $w_k$ and the measurement noise $v_k$ are modeled as zero-mean Gaussian with covariances $Q_k \in \mathbb{R}^{m \times m}$ and $R_k \in \mathbb{R}^{q \times q}$, as shown in (\ref{eq:3_gaussian_white}).

{\footnotesize
\begin{equation} \label{eq:3_gaussian_white}
\begin{aligned}
    &\mathbb{E} [w_k] = 0, \quad \mathbb{E} [v_k] = 0, \quad \mathbb{E} [w_i v_j] = 0 \\
    &\mathbb{E} [w_i w_j] = Q_k, \quad \mathbb{E} [v_i v_j] = R_k, \quad \forall i,\, j \in \mathbb{N} 
\end{aligned}
\end{equation} 
}

When used with MPCG, the AEKF employs the state vector $\bm{s}$ to avoid ambiguity, while the MPC formulation uses $\bm{x}$. The AEKF state vector, input, and measurement are defined in (\ref{eq:3_kf_states}), where the state includes relative position vector, velocity, target acceleration, and jerk in the navigation frame, the input is the missile acceleration, and the measurements are the relative range, its rate, and pitch/yaw look angles. The measurement model is given in (\ref{eq:3_meas}), where the third and fourth elements correspond to the pitch and yaw look angles components of the relative position vector, represented in (\ref{eq:2_look_angle}).

{\footnotesize
\begin{equation}  \label{eq:3_kf_states}
\begin{aligned}
    x &= \begin{bmatrix}
        \bm{r}^n, \, \bm{v}^n, \, \bm{a}^n_T, \, \bm{j}^n_T
    \end{bmatrix}^{\top}  \\
    u &= \begin{bmatrix}
        \bm{a}^n_M
    \end{bmatrix}^{\top} \\
    z &= \begin{bmatrix}
        R, \, \dot{R}, \, \sigma_\theta, \, \sigma_\psi
    \end{bmatrix}^{\top} 
\end{aligned}
\end{equation} 
}


{\footnotesize
\begin{align} \label{eq:3_meas}
\text{From (\ref{eq:2_rb_vb_def}),} \quad
        h_k (x) = 
    \begin{bmatrix}
         \sqrt{r^2_1 + r^2_2 + r^2_3} \\[0.5ex]
         \frac{r_1 \dot{r}_1 + r_2 \dot{r}_2 + r_3 \dot{r}_3}{\sqrt{r^2_1 + r^2_2 + r^2_3}} \\[0.5ex]
         \operatorname{atan} \left( \frac{-r_3}{\sqrt{r^2_1 + r^2_2}} \right) \\[0.5ex]
         \operatorname{atan2} \left( \frac{r_2}{r_1} \right)
    \end{bmatrix}
\end{align}
}

The Kalman filter algorithm generally consists of two steps: a prediction step, which propagates the state estimate from the previous time, and a correction step, which updates the prediction using measurement innovations. The process is summarized in the following equations.

{\footnotesize
\vspace{0.5ex}
\noindent \rule{\linewidth}{1.0pt}
\noindent \textbf{Algorithm 2:} \\ 
AEKF Algorithm \\ 
\noindent \rule{\columnwidth}{0.5pt}
\noindent \textbf{1. Initialization:}
\begin{align*}
\hat{x}_{0|0} &= \mathbb{E} \left[ x_0 \right] \\
{P}_{0|0} &= \mathbb{E} \left[ (x_0 - \hat{x}_{0|0}) (x_0 - \hat{x}_{0|0})^{\top} \right] 
\end{align*}

\vspace{1ex}
\noindent \textbf{2. Prediction:} 
\begin{align*}
    \hat{x}_{k|k-1} &= F_{k-1} \hat{x}_{k-1|k-1} + G_{k-1} u_{k-1} \\
    P_{k|k-1} &= F_{k-1} P_{k-1|k-1} F^{\top}_{k-1} + Q_{k-1}
\end{align*}

\vspace{1ex}
\noindent \textbf{3. Correction:}
\begin{align*}
    \hat{z}_{k|k-1} &= h_k (\hat{x}_{k|k-1}) \\
    \hat{H}_k &= \left. \frac{\partial h}{\partial x} \right|_{x_k=\hat{x}_{k|k-1}}
\end{align*}

\begin{align*}
    S_k &= \hat{H}_k P_{k|k-1} \hat{H}^{\top}_k + R_k \\
    K_k &= P_{k|k-1} \hat{H}^{\top}_k S^{-1}_k \\[0.5ex]
    \hat{x}_{k|k} &= \hat{x}_{k|k-1} + K_k (z_k - \hat{z}_{k|k-1}) \\
    P_{k|k} &= (I - K_k \hat{H}_k) P_{k|k-1} 
\end{align*}

\vspace{1ex}
\noindent \textbf{4. Update the $Q_{k-1}$ with innovation:}
\begin{align*}
    \hat{z}_{k|k-1} &= h_k(\hat{x}_{k|k-1}) \\
    d_k &= z_k - \hat{z}_{k|k-1} \\[1.5ex]
    w_{k-1} &= x_k - \left( F_{k-1} \hat{x}_{k-1|k-1} + G_{k-1} u_{k-1} \right) \\
    \hat{w}_{k-1} &= \hat{x}_{k|k} - \hat{x}_{k|k-1} \\
    &= K_k (z_k - \hat{z}_{k|k-1}) \\
    &= K_k d_k \\[1.5ex]
    \hat{Q}_{k-1} &= \mathbb{E}[\hat{w}_{k-1} \hat{w}_{k-1}^{\top}] \\
    &= \mathbb{E}[K_k (d_k d_k^{\top}) K_k^{\top}] \\
    &= K_k \mathbb{E}[d_k d_k^{\top}] K_k^{\top}
\end{align*}

\[
    Q_k = \alpha Q_{k-1} + (1 - \alpha) \hat{Q}_{k-1}
\]

\vspace{1ex}
\noindent \textbf{5. Update the $R_k$ with residual:}
\begin{align*}
    \hat{z}_{k|k} &= h_k(\hat{x}_{k|k}) \\
    \hat{H}_{k|k} &= \left. \frac{\partial h}{\partial x} \right|_{x_k = \hat{x}_{k|k}} \\
    \varepsilon_k &= z_k - \hat{z}_{k|k}
\end{align*}

\begin{align*}
    \hat{S}_k &= \mathbb{E}[\varepsilon_k \varepsilon_k^\top] \\
        &= \mathbb{E}[v_k v_k^\top] - \hat{H}_{k|k} P_{k|k} \hat{H}_{k|k}^\top \\
    \hat{R}_k &= \mathbb{E}[\varepsilon_k \varepsilon_k^\top] + \hat{H}_{k|k} P_{k|k} \hat{H}_{k|k}^\top
\end{align*}

\[
    R_{k+1} = \alpha R_k + (1 - \alpha) \mathbb{E}[\varepsilon_k \varepsilon_k^\top] + \hat{H}_{k|k} P_{k|k} \hat{H}_{k|k}^{\top}
\]

\noindent \rule{\columnwidth}{0.5pt}
}

In the Kalman filter, the posteriori estimate is denoted by the subscript $k|k$, while the priori estimate is denoted by $k|k-1$. The posteriori covariance $P_{k|k}$ is updated after incorporating the measurement, whereas the priori covariance $P_{k|k-1}$ is propagated from the previous step using the state dynamics and process noise.

To handle nonlinear measurement models, the EKF linearizes the function
$h(\cdot)$ around the priori estimate $\hat{x}_{k|k-1}$ using the Jacobian $\hat{H}_k = \left. \frac{\partial h}{\partial x} \right|_{x_k=\hat{x}_{k|k-1}}$. This allows the filter to approximate nonlinear measurements with a linear form, enabling the update to proceed in the same framework as the Linear Kalman Filter.

The innovation $d_k = z_k - \hat{z}_{k-1}$ and its covariance $S_k$ determine the Kalman gain $K_k$, which balances the relative confidence between the measurement and the model prediction. A smaller measurement noise covariance $R_k$ results in a larger Kalman gain, strengthening the influence of the measurement, whereas a smaller prior covariance $P_{k|k-1}$ reduces the gain, causing the filter to rely more on the prediction. The residual $\varepsilon_k = z_k - \hat{z}_{k}$, by contrast, quantifies the error with respect to the posteriori estimate.

While the EKF relies on fixed noise covariances, the AEKF enhances adaptability by updating $Q_{k-1}$ and $R_{k}$ at each step \cite{c17, c29, c30, c31, c32}. These updates are implemented through innovation- or residual-based formulations, often with a forgetting factor $\alpha \in (0,1)$ to balance past and recent information. In this way, the AEKF provides more reliable state estimation under time-varying noise conditions and abrupt target maneuvers, which are common in missile guidance scenarios.


\subsection{Interacting Multiple Model}

In practical missile–target engagement scenarios, the acceleration of a maneuvering target cannot be directly measured and must instead be estimated. Since the maneuvering intensity is generally unknown, Multiple Model (MM) Kalman filtering is employed, where several candidate models are run in parallel to represent different maneuvering levels. The use of MM structures in Kalman filter-based designs for target tracking or input estimation has been explored in prior studies \cite{c33, c34, c35, c36, c22, c37}.

\begin{figure}[!htbp]
  \centering
  \includegraphics[width=\columnwidth]{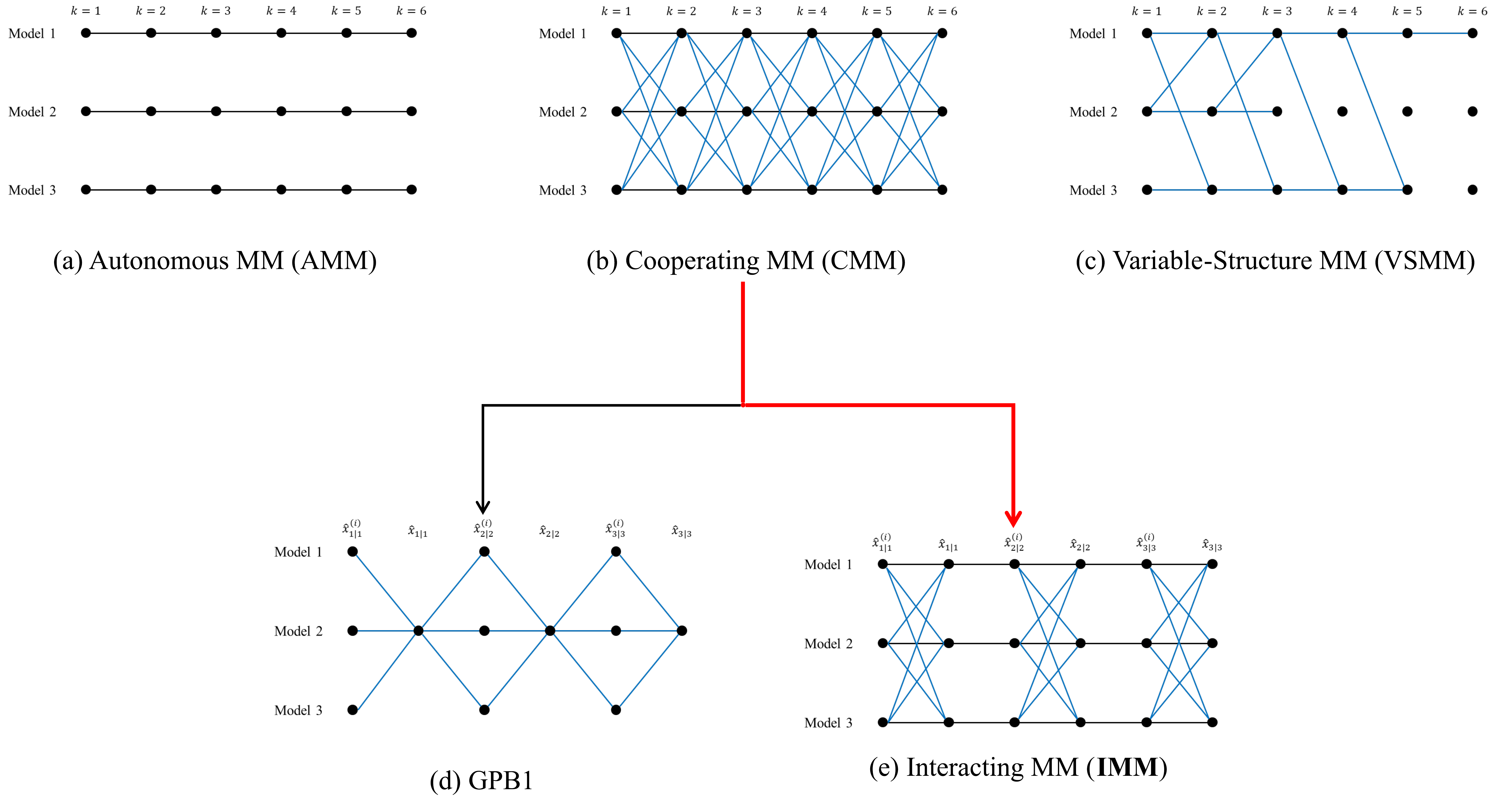}
  \caption{Classification of the multiple model Kalman filter}
  \label{fig:3_1_MM}
\end{figure}

As shown in Fig. \ref{fig:3_1_MM}, the MM framework consists of four core components: model-set determination, cooperation strategy, conditional filtering, and output processing. Depending on the structural configuration, MM approaches can be classified into three categories: Autonomous MM (AMM), Cooperating MM (CMM), and Variable-Structure MM (VSMM). AMM operates independent models without interaction, while CMM incorporates cooperation among models for improved estimation accuracy. VSMM allows dynamic changes of the model configuration depending on the system state or constraints.

Among these, the CMM structure is particularly suitable for online applications due to its balance between computational efficiency and estimation accuracy. Representative CMM algorithms include the Generalized Pseudo Bayesian 1 (GPB1) and the IMM. GPB1 accounts for model transition probabilities but maintains independent estimations, whereas IMM introduces interaction among models by mixing state information at each step.

In the IMM algorithm, multiple dynamic models operate concurrently, with each model generating its own state estimate. These estimates are then combined with the corresponding mode transition probabilities through an interactive reweighting process, which reflects the likelihood of switching between models. This mechanism produces a unified final state estimate that incorporates both the diversity of model predictions and the probabilistic transitions among them. By explicitly embedding mode transition probabilities into the estimation process, IMM effectively captures the time-varying dynamics of maneuvering targets. Consequently, it achieves more flexible and accurate tracking performance, particularly for nonlinear or highly maneuvering scenarios.

\begin{figure}[!htbp]
  \centering
  \includegraphics[width=\columnwidth]{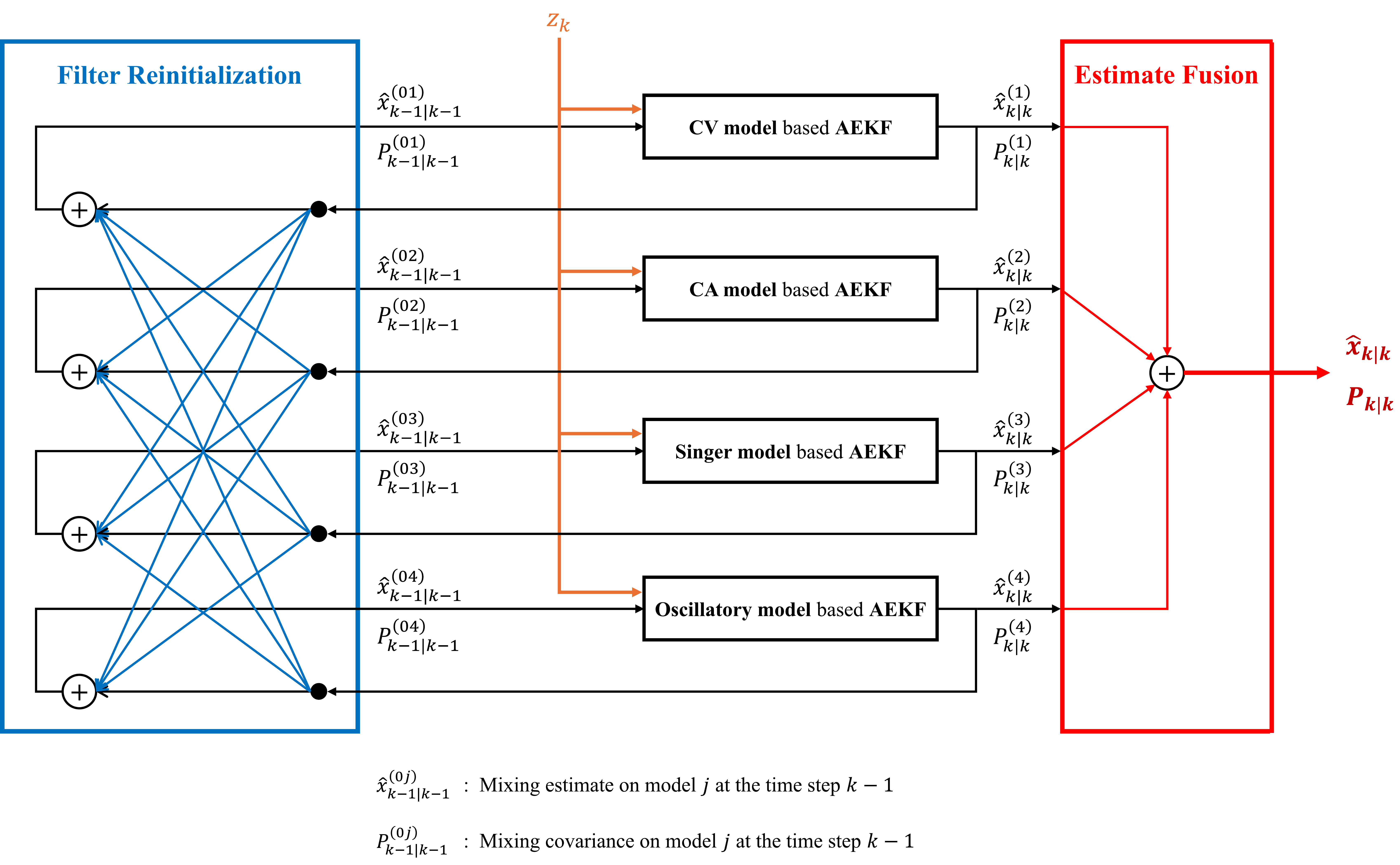}
  \caption{Structure of the IMM algorithm for target input estimation \cite{c22}}
  \label{fig:3_3_modeling_IMM}
\end{figure}

Fig. \ref{fig:3_3_modeling_IMM} illustrates the overall procedure of the IMM algorithm, which consists of four steps: model-conditioned reinitialization, model-conditioned filtering, model probability update, and estimate fusion.

The underlying MM framework assumes that the system dynamics can be represented by a set of $N$ models $M_j, ; j=1,\dots,N$, as expressed by

{\footnotesize
\begin{equation} \label{eq:mm_dyn}
\begin{aligned} 
    x_{k+1} = F_k^{(j)} x_k + G_k^{(j)} u_k + w_k^{(j)}, \quad
    z_k = h_k^{(j)}(x_k) + v_k^{(j)}
\end{aligned}
\end{equation}
}

At time $k$, the model probability of selecting model $M_j$ given all past measurements $Z_k$ is denoted as

{\footnotesize
\begin{equation} \label{eq:mm_prob}
    \omega^{(j)}_k = P \{ \sigma^{(j)}_k | Z_k \}
\end{equation}
}

where $\sigma^{(j)}_k$ indicates the active model. Model transitions are modeled as a Markov process, with the transition probability from model $M_i$ to $M_j$ denoted by $\pi_{ij}$.

{\footnotesize
\begin{equation} \label{eq:mm_trans}
    \pi_{ij} = P \{ \sigma^{(j)}_k | \sigma^{(i)}_{k-1} \}, \quad
    \pi_{ij} \geq 0, \quad \sum^N_{j=1} \pi_{ij} = 1 
\end{equation}
}

The set of $\pi_{ij}$ defines the model transition probability matrix $\Pi \in \mathbb{R}^{N \times N}$. Each model $M_j$ maintains its own state estimate and covariance independently, expressed as

{\footnotesize 
\begin{equation} \label{eq:mm_gauss}
    p( x_{k-1} | \sigma^{ (j) }_k, \sigma^{ (i) }_{k-1}, Z_{k-1} )
    = \mathcal{N} ( x_{k-1} | \hat{x}^{(i)}_{k-1|k-1}, P^{(i)}_{k-1|k-1} )
\end{equation}
}

Based on these properties, the IMM algorithm mixes prior estimates according to (\ref{eq:mm_trans}), applies independent filtering for each model, updates the mode probabilities via (\ref{eq:mm_prob}), and fuses the individual estimates into the final state. The IMM algorithm can be formulated as follows.

{\footnotesize
\vspace{0.5ex}
\noindent \rule{\linewidth}{1.0pt}
\noindent \textbf{Algorithm 3:} \\ 
AEKF-IMM Algorithm \\
\noindent \rule{\columnwidth}{0.5pt}
\noindent
\textbf{1. Model-conditioned reinitialization $\{ M_j, \quad j = 1, \dots, N \}:$} \\

\textit{Mixing probability update:}

\begin{align*}
    \mu_{k-1|k-1}^{(i|j)} = \frac{\pi_{ij} \omega_{k-1}^{(i)}}{\sum_{i=1}^{N} \pi_{ij} \omega_{k-1}^{(i)}}, \quad i = 1, \dots, N
\end{align*}

\textit{Mixing estimate and covariance:}

\begin{align*}
\hat{x}_{k-1|k-1}^{(0j)} &= \sum_{i=1}^{N} \mu_{k-1|k-1}^{(i|j)} \hat{x}_{k-1|k-1}^{(i)} \\[1ex]
e_0 &= \hat{x}_{k-1|k-1}^{(i)} - \hat{x}_{k-1|k-1}^{(0j)} \\[1ex]
P_{k-1|k-1}^{(0j)} &= \sum_{i=1}^{N} \mu_{k-1|k-1}^{(i|j)} \left[ P_{k-1|k-1}^{(i)} + e_0 e_0^\top \right]
\end{align*}

\vspace{1ex}
\noindent \textbf{2. Model-conditioned filtering $\{ M_j, \quad j = 1, \dots, N \}$:} \\ 

\textit{Prediction:}

\begin{align*}
\hat{x}^{(j)}_{k|k-1} &= F^{(j)}_{k-1} \hat{x}^{(0j)}_{k-1|k-1} + G^{(j)}_{k-1} u^{(j)}_{k-1} \\
P^{(j)}_{k|k-1} &= F^{(j)}_{k-1} P^{(0j)}_{k-1|k-1} \left( F^{(j)}_{k-1} \right)^{\top} + Q^{(j)}_{k-1}
\end{align*}

\textit{Correction:}

\begin{align*}
S^{(j)}_k &= \hat{H}^{(j)}_k P^{(j)}_{k|k-1} \left( \hat{H}^{(j)}_k \right)^{\top} + R^{(j)}_k \\
K^{(j)}_k &= P^{(j)}_{k|k-1} \left( \hat{H}^{(j)}_k \right)^{\top} \left( S^{(j)}_k \right)^{-1} \\
\hat{x}^{(j)}_{k|k} &= \hat{x}^{(j)}_{k|k-1} + K^{(j)}_k \left( z_k - \hat{z}^{(j)}_{k|k-1} \right) \\
P^{(j)}_{k|k} &= \left( I - K^{(j)}_k \hat{H}^{(j)}_k \right) P^{(j)}_{k|k-1}
\end{align*}
\\
\textit{Adaptive Noise Covariance:}

\begin{align*}
Q^{(j)}_k &= \alpha Q^{(j)}_{k-1} + (1 - \alpha) \hat{Q}^{(j)}_{k-1} \\
R^{(j)}_{k+1} &= \alpha R^{(j)}_k + (1 - \alpha) \mathbb{E} \left[ \varepsilon^{(j)}_k \left( \varepsilon^{(j)}_k \right)^{\top} \right] + \hat{H}^{(j)}_{k|k} P^{(j)}_{k|k} \left( \hat{H}^{(j)}_{k|k} \right)^{\top}
\end{align*}

\vspace{1ex}
\noindent \textbf{3. Model probability update $\{ M_j, \quad j = 1, \dots, N \}$:} \\ 

\textit{Model likelihood:}

\begin{align*}
    L^{(j)}_k = \mathcal{N} \left( z_k \mid \hat{z}^{(j)}_{k|k-1}, \; S^{(j)}_{k|k-1} \right)
\end{align*}

\textit{Model probability:}

\begin{align*}
    \omega^{(j)}_k = 
    \frac{L^{(j)}_k \sum_{i=1}^{N} \pi_{ij} \, \omega^{(i)}_{k-1}}{\sum_{j=1}^{N} L^{(j)}_k \sum_{i=1}^{N} \pi_{ij} \, \omega^{(i)}_{k-1}}
\end{align*}

\vspace{1ex}
\noindent \textbf{4. Estimation fusion:} 

\textit{Overall estimate and covariance:}
\begin{align*}
    \hat{x}_{k|k} &= \sum_{j=1}^{N} \omega^{(j)}_k \, \hat{x}^{(j)}_{k|k} \\[1ex]
    e_1 &= \hat{x}^{(j)}_{k|k} - \hat{x}_{k|k} \\[1ex]
    P_{k|k} &= \sum_{j=1}^{N} \omega^{(j)}_k \left[ P^{(j)}_{k|k} + e_1 e_1^\top \right]
\end{align*}

\noindent \rule{\columnwidth}{0.5pt}
}

The AEKF-IMM algorithm integrates adaptive estimation with multiple model interactions through four key steps. First, model-conditioned reinitialization mixes the state and covariance estimates according to the transition probabilities. Second, each model performs state estimation via the AEKF, including adaptive updates of the process and measurement noise covariances. Third, model probabilities are updated based on the likelihood of the measurements under each model. Finally, the individual state estimates are fused into an overall estimate weighted by the updated model probabilities. This procedure enables robust and adaptive tracking of maneuvering targets under uncertain dynamics.


\subsection{Modeling of Target Maneuvers within the IMM Framework}

Within the IMM framework, the maneuvering behavior of the target is represented through multiple motion models, each capturing a distinct dynamic characteristic. In this study, four motion models are considered: Constant Velocity (CV), Constant Acceleration (CA), Singer, and a Markov oscillatory model. Their dynamic matrices are derived in the continuous-time domain and then discretized for implementation, while a common seeker measurement model is applied across all modes as in (\ref{eq:3_rh_rx}).

From (\ref{eq:2_rb_vb_def}), (\ref{eq:2_R_R2}), (\ref{eq:3_kf_states}) and (\ref{eq:3_meas}),

\begin{align}
     R_\Sigma &= r_1 \dot{r}_1 + r_2 \dot{r}_2 + r_3 \dot{r}_3\\
    \frac{\partial h}{\partial \bm{x}} 
    &= \begin{bmatrix}
         \frac{\partial h}{\partial \bm{r}^n} & 
         \frac{\partial h}{\partial \bm{v}^n} & 
        \bm{0}_{4 \times 3} & \bm{0}_{4 \times 3}
    \end{bmatrix} \nonumber \\
    &= \begin{bmatrix}
         \frac{\partial h}{\partial \bm{r}^b} \frac{\partial \bm{r}^b}{\partial \bm{r}^n} &
         \frac{\partial h}{\partial \bm{v}^b} \frac{\partial \bm{v}^b}{\partial \bm{v}^n} &
        \bm{0}_{4 \times 3} & \bm{0}_{4 \times 3}
    \end{bmatrix} \label{eq:3_rh_rx} \\
    &= \begin{bmatrix}
         \frac{\partial h}{\partial \bm{r}^b} C_n^b &
         \frac{\partial h}{\partial \bm{v}^b} C_n^b &
        \bm{0}_{4 \times 3} & \bm{0}_{4 \times 3}
    \end{bmatrix} \nonumber \\[1ex]
    \frac{\partial h}{\partial \bm{r}^b} &=
    \begin{bmatrix}
         \frac{r_1}{R} &
         \frac{r_2}{R} &
         \frac{r_3}{R} \\
         \frac{\dot{r}_1}{R} - \frac{r_1 R_\Sigma}{R^3} &
         \frac{\dot{r}_2}{R} - \frac{r_2 R_\Sigma}{R^3} &
         \frac{\dot{r}_3}{R} - \frac{r_3 R_\Sigma}{R^3} \\
         \frac{r_1 r_3}{R^2 R_\perp} &  \frac{r_2 r_3}{R^2 R_\perp} &
         -\frac{R_\perp}{R^2} \\
         -\frac{r_2}{R_\perp^2} &
         \frac{r_1}{R_\perp^2} & 0
    \end{bmatrix} \\
    \frac{\partial h}{\partial \bm{v}^b} &=
    \begin{bmatrix}
        0 & 0 & 0 \\
         \frac{r_1}{R} &
         \frac{r_2}{R} &
         \frac{r_3}{R} \\
        0 & 0 & 0 \\
        0 & 0 & 0
    \end{bmatrix}
\end{align}

The subsequent subsections provide the detailed formulations and design aspects of the four maneuver models: CV, CA, Singer, and a Markov model for oscillatory maneuvers.

\subsubsection{Constant Velocity}  %
If the target is assumed to move with constant velocity in three-dimensional space, its continuous-time dynamics can be expressed as

\begin{equation}
    \begin{aligned}
        \dot{\bm{v}}^n &= \bm{a}_T^n - \bm{a}_M^n + w(t) \\
\dot{{x}}(t) &= A_{CV} x(t) + B_{CV} u(t) + Q_c w(t) \\
    \end{aligned}
\end{equation}

{\footnotesize
\begin{equation} \label{eq:3_cv_cont}
    \begin{aligned}
    \bm{0}_{3} &= \bm{0}_{3 \times 3}, \quad \bm{I}_{3} = \bm{I}_{3 \times 3}  \\
        \dot{x}(t) &=
        \begin{bmatrix}
        \bm{0}_{3} & \bm{I}_{3} & \bm{0}_{3} & \bm{0}_{3} \\
        \bm{0}_{3} & \bm{0}_{3} & \bm{I}_{3} & \bm{0}_{3} \\
        \bm{0}_{3} & \bm{0}_{3} & \bm{0}_{3} & \bm{I}_{3} \\
        \bm{0}_{3} & \bm{0}_{3} & \bm{0}_{3} & \bm{0}_{3}
        \end{bmatrix}
        x(t) 
        +
        \begin{bmatrix}
        \bm{0}_{3} \\
        - \bm{I}_{3} \\
        \bm{0}_{3} \\
        \bm{0}_{3}
        \end{bmatrix}
        u(t) +
        \begin{bmatrix}
        \bm{0}_{3} \\
        \bm{I}_{3} \\
        \bm{0}_{3} \\
        \bm{0}_{3}
        \end{bmatrix}
        w(t)
    \end{aligned} 
\end{equation}
}

where, $w(t) \sim \mathcal{N}(0, \, Q_c)$ denotes Gaussian white noise accounting for external disturbances. The process noise covariance, which corresponds to the power spectral density (PSD) of the noise, is 
{\footnotesize
\[
Q_c =
\begin{bmatrix}
\sigma_{x,CV}^2 & 0 & 0 \\
0 & \sigma_{y,CV}^2 & 0 \\
0 & 0 & \sigma_{z,CV}^2
\end{bmatrix}
\]}.

This noise, which reflects residual accelerations, determines the maneuver intensity; hence the model is also referred to as the nearly constant velocity (NCV) model.

The continuous-time linear system can be converted into an equivalent discrete time model using the matrix exponential of the state transition matrix \cite{c38}. Discretizing (\ref{eq:3_cv_cont}) with sampling time $\Delta t$, the discrete-time model becomes

\begin{align}
x_{k+1} &= F_{CV} x_k + G_{CV} u_k + w_k  \\
x_{k+1} &=
\begin{bmatrix}
\bm{I}_{3} & \Delta t\, \bm{I}_{3} & \frac{(\Delta t)^2}{2} \bm{I}_{3} & \frac{(\Delta t)^3}{6} \bm{I}_{3} \\
\bm{0}_{3} & \bm{I}_{3} & \Delta t\, \bm{I}_{3} & \frac{(\Delta t)^2}{2} \bm{I}_{3} \\
\bm{0}_{3} & \bm{0}_{3} & \bm{I}_{3} & \Delta t\, \bm{I}_{3} \\
\bm{0}_{3} & \bm{0}_{3} & \bm{0}_{3} & \bm{I}_{3}
\end{bmatrix}
x_k \\  &+
\begin{bmatrix}
- \frac{(\Delta t)^2}{2} \bm{I}_{3} \\
- \Delta t\, \bm{I}_{3} \\
\bm{0}_{3} \\
\bm{0}_{3}
\end{bmatrix}
u_k + w_k \nonumber
\end{align}

with $w_k \sim \mathcal{N}(0, Q_k)$, and

\begin{equation}
Q_k = \sigma_{CV}^2
\begin{bmatrix}
\frac{(\Delta t)^3}{3} \bm{I}_{3} & \frac{(\Delta t)^2}{2} \bm{I}_{3} & \bm{0}_{3} & \bm{0}_{3} \\
\frac{(\Delta t)^2}{2} \bm{I}_{3} & \Delta t\, \bm{I}_{3} & \bm{0}_{3} & \bm{0}_{3} \\
\bm{0}_{3} & \bm{0}_{3} & \bm{0}_{3} & \bm{0}_{3} \\
\bm{0}_{3} & \bm{0}_{3} & \bm{0}_{3} & \bm{0}_{3} 
\end{bmatrix} 
\end{equation}

\subsubsection{Constant Acceleration}

If the target is assumed to follow a constant acceleration motion in three-dimensional space, its dynamics can be expressed as follows.

\begin{equation}
    \begin{aligned}
        \dot{\bm{a}}_T^n &= w(t) \\
        \dot{x}(t) &= A_{CA} x(t) + B_{CA} u(t) + Q_c w(t)
    \end{aligned}
\end{equation}

{\footnotesize
\begin{equation}
    \begin{aligned}
        \dot{x}(t) &=
        \begin{bmatrix}
        \bm{0}_{3} & \bm{I}_{3} & \bm{0}_{3} & \bm{0}_{3} \\
        \bm{0}_{3} & \bm{0}_{3} & \bm{I}_{3} & \bm{0}_{3} \\
        \bm{0}_{3} & \bm{0}_{3} & \bm{0}_{3} & \bm{I}_{3} \\
        \bm{0}_{3} & \bm{0}_{3} & \bm{0}_{3} & \bm{0}_{3}
        \end{bmatrix}
        x(t) +
        \begin{bmatrix}
        \bm{0}_{3} \\
        - \bm{I}_{3} \\
        \bm{0}_{3} \\
        \bm{0}_{3}
        \end{bmatrix}
        u(t) +
        \begin{bmatrix}
        \bm{0}_{3} \\
        \bm{0}_{3} \\
        \bm{I}_{3} \\
        \bm{0}_{3}
        \end{bmatrix}
        w(t)
    \end{aligned}
\end{equation}
}

Compared to the CV model represented in (\ref{eq:3_cv_cont}), the main difference lies in the presence of an identity matrix in the third block of the $Q_c$ matrix. Since the power spectral density
{\footnotesize
\[
Q_c =
\begin{bmatrix}
\sigma_{x,CA}^2 & 0 & 0 \\
0 & \sigma_{y,CA}^2 & 0 \\
0 & 0 & \sigma_{z,CA}^2
\end{bmatrix}
\]
}
induces jerk motion of the target, this model is often referred to as the nearly constant acceleration model.

The continuous-time system is discretized using the sampling interval $\Delta t$, resulting in the following form.

\begin{align}
x_{k+1} &= F_{CA} x_k + G_{CA} u_k + w_k  \\
x_{k+1} &= 
\begin{bmatrix}
\bm{I}_{3} & \Delta t\, \bm{I}_{3} & \frac{(\Delta t)^2}{2} \bm{I}_{3} & \frac{(\Delta t)^3}{6} \bm{I}_{3} \\
\bm{0}_{3} & \bm{I}_{3} & \Delta t\, \bm{I}_{3} & \frac{(\Delta t)^2}{2} \bm{I}_{3} \\
\bm{0}_{3} & \bm{0}_{3} & \bm{I}_{3} & \Delta t\, \bm{I}_{3} \\
\bm{0}_{3} & \bm{0}_{3} & \bm{0}_{3} & \bm{I}_{3}
\end{bmatrix}
x_k \\ 
&+
\begin{bmatrix}
- \frac{(\Delta t)^2}{2} \bm{I}_{3} \\
- \Delta t\, \bm{I}_{3} \\
\bm{0}_{3} \\
\bm{0}_{3}
\end{bmatrix}
u_k + w_k \nonumber
\end{align}

In this context, $w_k$ denotes the discretized process noise, characterized by the following mean and covariance.

\begin{equation}
\begin{aligned}
w_k &\sim \mathcal{N}(0, Q_k) \\
Q_k &= \sigma_{CA}^2
\begin{bmatrix}
\frac{(\Delta t)^5}{20} \bm{I}_{3} & \frac{(\Delta t)^4}{8} \bm{I}_{3} & \frac{(\Delta t)^3}{6} \bm{I}_{3} & \bm{0}_{3} \\
\frac{(\Delta t)^4}{8} \bm{I}_{3} & \frac{(\Delta t)^3}{3} \bm{I}_{3} & \frac{(\Delta t)^2}{2} \bm{I}_{3} & \bm{0}_{3} \\
\frac{(\Delta t)^3}{6} \bm{I}_{3} & \frac{(\Delta t)^2}{2} \bm{I}_{3} & \Delta t\, \bm{I}_{3} & \bm{0}_{3} \\
\bm{0}_{3} & \bm{0}_{3} & \bm{0}_{3} & \bm{0}_{3}
\end{bmatrix} 
\end{aligned}
\end{equation}

\subsubsection{Singer Acceleration Model}

The Singer model assumes that target acceleration follows a zero-mean first-order stationary Markov stochastic process, where the correlation time constant $\tau$ controls the persistence and maneuverability of the acceleration. As $\tau$ increases, the model approximates a CV behavior, while a smaller $\tau$ causes it to resemble a CA model.

\begin{equation}
    \dot{\bm{a}}^n_T = -\alpha \bm{a}^n_T + w (t)
\end{equation}

By simultaneously incorporating both temporal correlation and process noise, the Singer model captures more realistic maneuvering behaviors. It also allows for flexible modeling of acceleration variance using probabilistic distributions such as ternary-uniform mixtures.

The singer model dynamics can be expressed as follows.

{\footnotesize
\begin{align}
    \dot{x}(t) &= {A}_{SR} x(t) + {B}_{SR} u(t) + {Q}_c w(t) \\
    \dot{x}(t) &=
    \begin{bmatrix}
    \bm{0}_{3} & \bm{I}_{3} & \bm{0}_{3} & \bm{0}_{3} \\
    \bm{0}_{3} & \bm{0}_{3} & \bm{I}_{3} & \bm{0}_{3} \\
    \bm{0}_{3} & \bm{0}_{3} & -\alpha \bm{I}_{3} & \bm{I}_{3} \\
    \bm{0}_{3} & \bm{0}_{3} & \bm{0}_{3} & \bm{0}_{3}
    \end{bmatrix}
    x(t)  +
    \begin{bmatrix}
    \bm{0}_{3} \\
    -\bm{I}_{3} \\
    \bm{0}_{3} \\
    \bm{0}_{3}
    \end{bmatrix}
    u(t)
    +
    \begin{bmatrix}
    \bm{0}_{3} \\
    \bm{0}_{3} \\
    \bm{I}_{3} \\
    \bm{0}_{3}
    \end{bmatrix}
    w(t) \nonumber
\end{align}
}

The discrete-time model is obtained as follows, using the sampling interval $\Delta t$,

\begin{align}
x_{k+1} &= F_{SR} x_k + G_{SR} u_k + w_k  \\[0.5ex]
x_{k+1} &=
\begin{bmatrix}
\bm{I}_{3} & \Delta t\, \bm{I}_{3} & (F_{SR})_{13} & (F_{SR})_{14} \\
\bm{0}_{3} & \bm{I}_{3} & (F_{SR})_{23} & (F_{SR})_{24} \\
\bm{0}_{3} & \bm{0}_{3} & (F_{SR})_{33} & (F_{SR})_{34}  \\
\bm{0}_{3} & \bm{0}_{3} & \bm{0}_{3} & \bm{I}_{3}
\end{bmatrix}
x_k \\
&+
\begin{bmatrix}
- \frac{(\Delta t)^2}{2} \bm{I}_{3} \\
- \Delta t\, \bm{I}_{3} \\
\bm{0}_{3} \\
\bm{0}_{3}
\end{bmatrix}
u_k + w_k \nonumber
\end{align}

{\footnotesize
\begin{align*}
(F_{SR})_{13} &= \frac{\alpha \Delta t - \left(1 - e^{-\alpha \Delta t} \right)}{\alpha^2} \bm{I}_{3} \\
(F_{SR})_{14} &= \frac{\alpha (\Delta t)^2 - 2\Delta t + 2\left(1 - e^{-\alpha \Delta t} \right)}{\alpha^3} \bm{I}_{3} \\
(F_{SR})_{23} &= \frac{1 - e^{-\alpha \Delta t}}{\alpha} \bm{I}_{3} \\
(F_{SR})_{24} &= \frac{\alpha \Delta t - \left(1 - e^{-\alpha \Delta t} \right)}{\alpha^2} \bm{I}_{3} \\
(F_{SR})_{33} &= e^{-\alpha \Delta t} \bm{I}_{3} \\
(F_{SR})_{34} &= \frac{1 - e^{-\alpha \Delta t}}{\alpha} \bm{I}_{3}
\end{align*}
}

The discretized process noise $w_k$ has the covariance matrix $Q_k$ as shown below \cite{c23}.

\begin{align}
    w_k &\sim \mathcal{N}(0, Q_k) \\
    Q_k &= \sigma^2_{SR}
    \begin{bmatrix}
    Q_{11} \bm{I}_{3} & Q_{12} \bm{I}_{3} & Q_{13} \bm{I}_{3} & \bm{0}_{3} \\
    Q_{12} \bm{I}_{3} & Q_{22} \bm{I}_{3} & Q_{23} \bm{I}_{3} & \bm{0}_{3} \\
    Q_{13} \bm{I}_{3} & Q_{23} \bm{I}_{3} & Q_{33} \bm{I}_{3} & \bm{0}_{3} \\
    \bm{0}_{3} & \bm{0}_{3} & \bm{0}_{3} & \bm{0}_{3}
    \end{bmatrix} \nonumber
\end{align}

\begin{align*}
Q_{11} &= \frac{1}{\alpha^4} \bigg[ 1 - e^{-2\alpha \Delta t} + 2\alpha \Delta t 
\\ &\qquad + \frac{2}{3} (\alpha \Delta t)^3
- 2(\alpha \Delta t)^2 - 4\alpha \Delta t e^{-\alpha \Delta t} \bigg] \\
Q_{12} &= \frac{1}{\alpha^3} \bigg[ e^{-2\alpha \Delta t} + 1 - 2e^{-\alpha \Delta t} 
\\ &\qquad + 2\alpha \Delta t e^{-\alpha \Delta t}
- 2\alpha \Delta t + (\alpha \Delta t)^2 \bigg] \\
Q_{13} &= \frac{1}{\alpha^2} \left[ -2\alpha \Delta t e^{-\alpha \Delta t} 
+ \left( 1 - e^{-2\alpha \Delta t} \right) \right] \\
Q_{22} &= \frac{1}{\alpha^2} \left[ 2\alpha \Delta t - 4\left(1 - e^{-\alpha \Delta t}\right) 
+ \left(1 - e^{-2\alpha \Delta t}\right) \right] \\
Q_{23} &= \frac{1}{\alpha} \left[ 2\left(1 - e^{-\alpha \Delta t}\right) 
- \left(1 - e^{-2\alpha \Delta t}\right) \right] \\
Q_{33} &= 1 - e^{-2\alpha \Delta t}
\end{align*}

If the sampling time is small, i.e., $\alpha \Delta t \ll \frac{1}{2}$, the noise covariance $Q_k$ may be used:

\begin{align} 
Q_k &\approx \bar{Q}_k  \label{eq:3_CA_Q} \\  
 &=  
2 \alpha \sigma^2_{SR} 
\begin{bmatrix} 
\frac{(\Delta t)^5}{20} \bm{I}_{3} & \frac{(\Delta t)^4}{8} \bm{I}_{3} & \frac{(\Delta t)^3}{6} \bm{I}_{3} & \bm{0}_{3} \\
\frac{(\Delta t)^4}{8} \bm{I}_{3} & \frac{(\Delta t)^3}{3} \bm{I}_{3} & \frac{(\Delta t)^2}{2} \bm{I}_{3} & \bm{0}_{3} \\
\frac{(\Delta t)^3}{6} \bm{I}_{3} & \frac{(\Delta t)^2}{2} \bm{I}_{3} & \Delta t\, \bm{I}_{3} & \bm{0}_{3} \\
\bm{0}_{3} & \bm{0}_{3} & \bm{0}_{3} & \bm{0}_{3}
\end{bmatrix} \nonumber
\end{align}

\subsubsection{Markov Models for Oscillatory Target}

The oscillatory target model is suitable for scenarios where the target's acceleration in a specific lateral direction exhibits periodic oscillations. Unlike the non-periodic nature of the Singer model, this model is characterized by a periodic autocorrelation function $R_a (\tau)$. It is represented as a \textit{second-order pre-whitening system} that produces target acceleration in response to zero-mean white noise inputs. The transfer function is given as

{\footnotesize
\begin{equation} \label{eq:3_transfer_func}
    H(s) = \frac{s + \omega_n}{s^2 + 2\zeta \omega_n s + \omega_n^2}, 
    \quad \omega_n^2 = \alpha^2 + \omega_c^2, 
    \quad \zeta = \frac{\alpha}{\omega_n}   
\end{equation}
}

Here $\zeta, \omega_n, \omega_c,$ and $\alpha$ denotes damping ratio, undamped natural frequency, actual (damped) frequency, and damping coefficient of the target acceleration, respectively. The system is governed by parameters such as damping ratio $\zeta$, natural frequency $\omega_n$. Due to its ability to capture high-frequency oscillatory behavior more effectively than the Singer model, this model is well suited for targets exhibiting characteristics such as bending motion.

The frequency-domain representation in (\ref{eq:3_transfer_func}) can be expressed in the time domain as follows.

{\footnotesize
\begin{equation} \label{eq:3_acceleration_drift}
\begin{bmatrix}
\dot{\bm{a}}(t) \\
\dot{\bm{d}}(t)
\end{bmatrix}
=
\begin{bmatrix}
0 & 1 \\
-\omega_n^2 & -2\zeta\omega_n
\end{bmatrix}
\begin{bmatrix}
\bm{a} (t) \\
\bm{d} (t)
\end{bmatrix}
+
\begin{bmatrix}
1 \\
(1 - 2\zeta)\omega_n
\end{bmatrix}
w(t)
\end{equation}
}

Here, $\dot{\bm{d}}(t)$ denotes the acceleration drift, defined as $\dot{\bm{a}}(t) - w(t)$. When acceleration drift is expressed in terms of jerk, which is a state variable in the Kalman filter, equation (\ref{eq:3_acceleration_drift}) is transformed as follows.

{\footnotesize
\begin{align}
    \bm{d} (t) 
        &= \dot{\bm{a}}(t) - w(t)  = \bm{j}(t) - w(t)  \\
    \frac{{d} \bm{j}(t)}{dt} - \dot{w}(t)
        &= 
    -\omega_n^2 \bm{a} (t) - 2\zeta \omega_n \big(\bm{j}(t) - w(t)\big) \label{eq:3_djdt} \\
    & \quad + \big[(1 - 2\zeta)\omega_n\big] w(t)  \nonumber \\
    \frac{d \bm{j} (t)}{dt}
        &\approx -\omega_n^2 \bm{a} (t) - 2\zeta \omega_n \bm{j} (t) + \omega_n w(t), 
        \quad \dot{w}(t) = 0 \nonumber \\
        \frac{d}{dt}
        \begin{bmatrix}
        \bm{a} (t) \\
        \bm{j} (t)
        \end{bmatrix}
        &=
        \begin{bmatrix}
        0 & 1 \\
        -\omega_n^2 & -2\zeta \omega_n
        \end{bmatrix}
        \begin{bmatrix}
        \bm{a} (t) \\
        \bm{j} (t)
        \end{bmatrix}
        +
        \begin{bmatrix}
        1 \\
        \omega_n
        \end{bmatrix}
        w(t)
\end{align}
}

In (\ref{eq:3_djdt}), the derivative of the process noise, $\dot{w}(t)$, is considered zero because it represents a non-differentiable process due to the inherent characteristics of white noise. The dynamics of the Markov models for Oscillatory target can be represented as follows.

\begin{align}
\dot{x}(t) &= A_{OS} x(t) + B_{OS} u(t) + Q_c w(t) \\
\dot{x}(t) &=
\begin{bmatrix}
\bm{0}_{3} & \bm{I}_{3} & \bm{0}_{3} & \bm{0}_{3} \\
\bm{0}_{3} & \bm{0}_{3} & \bm{I}_{3} & \bm{0}_{3} \\
\bm{0}_{3} & \bm{0}_{3} & \bm{0}_{3} & \bm{I}_{3} \\
\bm{0}_{3} & \bm{0}_{3} & -\omega_n^2 \bm{I}_{3} & -2\zeta \omega_n \bm{I}_{3}
\end{bmatrix}
x(t) +
\begin{bmatrix}
\bm{0}_{3} \\
-\bm{I}_{3} \\
\bm{0}_{3} \\
\bm{0}_{3}
\end{bmatrix}
u(t)
\\ &+
\begin{bmatrix}
\bm{0}_{3} \\
\bm{0}_{3} \\
\bm{I}_{3} \\
\omega_n \bm{I}_{3}
\end{bmatrix}
w(t) \nonumber 
\end{align}

The discrete-time model form is \cite{c33}.

\begin{align}
x_{k+1} &= F_{OS} x_k + G_{OS} u_k + w_k  \\
x_{k+1} &=
\begin{bmatrix}
\bm{I}_{3} & \Delta t\, \bm{I}_{3} & (F_{OS})_{13} & (F_{OS})_{14} \\
\bm{0}_{3} & \bm{I}_{3} & (F_{OS})_{23} & (F_{OS})_{24} \\
\bm{0}_{3} & \bm{0}_{3} & (F_{OS})_{33} & (F_{OS})_{34} \\
\bm{0}_{3} & \bm{0}_{3} & (F_{OS})_{43} & (F_{OS})_{44}
\end{bmatrix}
x_k 
\\ &+
\begin{bmatrix}
- \dfrac{(\Delta t)^2}{2} \bm{I}_{3} \\
- \Delta t\, \bm{I}_{3} \\
\bm{0}_{3} \\
\bm{0}_{3}
\end{bmatrix}
u_k + w_k \nonumber
\end{align}

{\footnotesize
\begin{align*}
swT &= \sin(\omega_c \Delta t), \quad cwT = \cos(\omega_c \Delta t) \\
(F_{OS})_{13}  &= \frac{1}{\omega_c(\alpha^2 + \omega_c^2)^2} \Big\{
\omega_c(-3\alpha^2 + \omega_c^2 + 2\alpha^3 \Delta t + 2\alpha \omega_c^2 \Delta t) \\
&\quad - e^{-\alpha \Delta t}
\left[\omega_c(-3\alpha^2 + \omega_c^2) cwT - \alpha(\alpha^2 - 3\omega_c^2) swT\right]
\Big\} \\
(F_{OS})_{14} &= \frac{1}{\omega_c (\alpha^2 + \omega_c^2)^2} \Big\{
\omega_c(-2\alpha + \alpha^2 \Delta t + \omega_c^2 \Delta t) \\
&\quad + e^{-\alpha \Delta t} \left[2\alpha \omega_c cwT + (\alpha^2 - \omega_c^2) swT \right]
\Big\} \\
(F_{OS})_{23} &= \frac{
2\alpha \omega_c - e^{-\alpha \Delta t}
\left[2\alpha \omega_c cwT + (\alpha^2 - \omega_c^2) swT \right]
}{
\omega_c (\alpha^2 + \omega_c^2)
} \\
(F_{OS})_{24} &= \frac{
\omega_c - e^{-\alpha \Delta t}
\left[\omega_c cwT + \alpha swT \right]
}{
\omega_c (\alpha^2 + \omega_c^2)
} \\
(F_{OS})_{33} &= \frac{
e^{-\alpha \Delta t} \left[\omega_c cwT + \alpha swT \right]
}{\omega_c} \\
(F_{OS})_{34} &= \frac{
e^{-\alpha \Delta t} swT
}{\omega_c} \\
(F_{OS})_{43} &= -\frac{
(\alpha^2 + \omega_c^2) e^{-\alpha \Delta t} swT
}{\omega_c} \\
(F_{OS})_{44} &= \frac{
e^{-\alpha \Delta t} \left[\omega_c cwT - \alpha swT \right]
}{\omega_c}
\end{align*}
}

Although the covariance, 
{\footnotesize
\[ cov(w_k) = S_w \int_0^{\Delta t} e^{A_\tau} BB^\top (e^{A_\tau})^\top  d\tau
\]}, can be derived using symbolic computation tools, the process is complicated and not presented here. Instead, a simplified discrete time noise covariance similar to (\ref{eq:3_CA_Q}) is used with $w_k \sim \mathcal{N}(0, Q_k)$.

\begin{align}
Q_k &\approx \bar{Q}_k \\
&= 2\alpha \sigma_{OS}^2
\begin{bmatrix}
\frac{(\Delta t)^7}{252}\,\bm{I}_3 & \frac{(\Delta t)^6}{72}\,\bm{I}_3 & \frac{(\Delta t)^5}{20}\,\bm{I}_3 & \frac{(\Delta t)^4}{8}\,\bm{I}_3 \\
\frac{(\Delta t)^6}{72}\,\bm{I}_3 & \frac{(\Delta t)^5}{20}\,\bm{I}_3 & \frac{(\Delta t)^4}{8}\,\bm{I}_3 & \frac{(\Delta t)^3}{6}\,\bm{I}_3 \\
\frac{(\Delta t)^5}{20}\,\bm{I}_3 & \frac{(\Delta t)^4}{8}\,\bm{I}_3 & \frac{(\Delta t)^3}{3}\,\bm{I}_3 & \frac{(\Delta t)^2}{2}\,\bm{I}_3 \\
\frac{(\Delta t)^4}{8}\,\bm{I}_3 & \frac{(\Delta t)^3}{6}\,\bm{I}_3 & \frac{(\Delta t)^2}{2}\,\bm{I}_3 & \Delta t\,\bm{I}_3
\end{bmatrix} \nonumber
\end{align}


\section{Numerical Simulation} \label{sec:4_numerical}

\subsection{Simulation Environment}

\subsubsection{Experimental Setup Descriptions}

To comprehensively evaluate the performance and accuracy of the proposed MPCG scheme, three representative target maneuver scenarios were designed: pitch-plane weaving, yaw-plane weaving, and barrel-roll. This configuration was selected to verify the algorithm’s adaptability across different engagements, including the vertical plane, lateral plane, and full 3D space. 

The pitch-plane weaving maneuver induces rapid vertical oscillations that temporarily force the target outside the missile's pitch look angle limit, thereby challenging the vertical guidance authority. The yaw-plane weaving maneuver consists of periodic side-to-side oscillations in the horizontal plane, forming an S-shaped path aimed at stressing the missile's lateral tracking capability.

The barrel-roll maneuver corresponds to circular motion perpendicular to the flight direction while maintaining forward velocity. Of particular interest is the high-G barrel-roll, executed at the target’s maximum allowable acceleration \cite{c39_0, c39, c40, c41, c42, c43}. In this maneuver, the minimum turning radius is governed by both the maximum acceleration and the angular roll rate $\omega_{br}$. The time-varying lateral accelerations in the body frame are modeled as:

\begin{equation}
    \begin{aligned}
        {a}_{2, T} &= a_{T, max} \cos{\omega_{br} t} \\
        {a}_{3, T} &= a_{T, max} \sin{\omega_{br} t}
    \end{aligned}
\end{equation}

Together, these scenarios serve to validate the MPCG’s ability to maintain effectiveness under diverse and highly dynamic target conditions.

\subsubsection{Guidance Implementation}

To implement the above scenario, the proposed MPCG algorithm was coded in Julia, with the continuous-time OCP discretized using LGRPM and solved as an NLP via JuMP.jl and IPOPT.jl, employing the MA97 sparse linear solver with a warm-start strategy to enhance computational efficiency \cite{c44, c45}.

To capture the autopilot dynamics of missile, a first-order lag system was adopted, yielding the following dynamics:

\begin{equation}
        \dot{\bm{a}}_{M} = \frac{ (\bm{a}_{M, cmd} - \bm{a}_{M})}{\tau_M}
\end{equation}

The time constants $\tau_T$ and $\tau_M$ characterize the system’s responsiveness to command inputs. For comparative analysis, a conventional PPNG was considered, and the estimated LOS angular velocity $\hat{\bm{\Omega}}$ was generated by adding Gaussian noise from the seeker to the true LOS angular velocity:

\begin{equation}
    \text{From (\ref{eq:2_LOS_angular_vel}),} \quad
    \hat{\bm{\Omega}} = \bm{\Omega}_{true} + e, \quad e \sim \mathcal{N} (0, \frac{0.001 \pi}{180})
\end{equation}

Using $\hat{\bm{\Omega}}$, the PPNG acceleration command was given as:

{\footnotesize
\begin{equation}
\begin{aligned}
    \bm{a}_{PPNG} &= N_{ppng} \hat{\bm{\Omega}} \times \bm{v}_M - \bm{g}_p, \quad
    \bm{g}_p = \bm{g} - (\bm{g} \cdot \hat{\bm{v}}_M) \hat{\bm{v}}_M 
\end{aligned}
\end{equation}
}

Where $\hat{\bm{v}}_M$ denotes the missile's unit velocity vector, and $\bm{g}_p$ compensates for gravitational effects along the missile’s flight path.

Furthermore, MD serves as a key performance metric in the design of missile guidance laws, defined as the minimum relative distance between the missile and the target during engagement. In the simulation, evasion was assumed under the conditions

\begin{equation}
    \| \bm{r} \| \leq  R_f \quad \text{and} \quad V_c = - \bm{v} \cdot \hat{\bm{r}} < 0 
\end{equation}

Here, $V_c$ represents the closing velocity between the missile and the target. the final MD is recorded as the relative distance at the last time step immediately before these conditions are met.

\subsubsection{Design Parameters}

The design parameters in both MPCG and AEKF-IMM can be categorized into three groups: guidance weighting matrices, estimator covariance matrices, and seeker constraints. For MPCG, the weighting matrices $\bm{Q}, \bm{Q}_f,$ and $\bm{R}$, as defined in (\ref{eq:2_mpcg_const_2}) regulate the balance between performance and stability. For AEKF-IMM, estimation accuracy is governed by the state error covariance $P$, the process noise covariance $Q$, and measurement noise covariance $R$. Finally, the strapdown seeker’s FOV, defined by its allowable look angle and detection range, imposes physical constraints that directly influence guidance evaluation. A complete summary of these parameters is provided in the following table.

{\scriptsize
\begin{table}[!htbp]
\caption{Summary of parameters by the framework component}
\centering
\begin{adjustbox}{max width=\columnwidth}
\begin{tabular}{@{}llll@{}} 
\toprule\toprule
\textbf{Component} & \textbf{Parameter} & \textbf{Value} & \textbf{Unit} \\
\midrule
\multirow{3}{*}{Missile}
  & $\Delta t$ & 0.01 & \text{sec} \\
  & $G$        & 9.80665 & m/s$^2$ \\
  & $R_{f}  $  & $10$    & m       \\
\midrule
\multirow{5}{*}{MPCG}
  & $\bm{Q}$   & $30 \cdot \mathrm{diag}([\bm{I}_{3 \times 3},\ 10^{-4} \bm{I}_{3 \times 3},\ \bm{I}_{2 \times 2},\ 10^{-4} \bm{I}_{2 \times 2},\ \bm{I}_{2 \times 2}])$ & -- \\
  & $\bm{Q}_f$ & $30 \cdot \mathrm{diag}([\bm{I}_{3 \times 3},\ 10^{-4} \bm{I}_{3 \times 3},\ \bm{I}_{2 \times 2},\ 10^{-4} \bm{I}_{2 \times 2},\ \bm{I}_{2 \times 2}])$ & -- \\
  & $\bm{R}$   & $15 I_{3 \times 3}$ & -- \\
  & $N$        & 7 & -- \\
  & $t_{\text{horizon}}$ & 0.5 & sec \\
\midrule
\multirow{7}{*}{AEKF-IMM}
  & $P_0$      & $\mathrm{diag}([100^2 \bm{I}_{3 \times 3},\ 20^2 \bm{I}_{3 \times 3},\ (2.5 G)^2 \bm{I}_{3 \times 3},\ (1.5 G)^2 \bm{I}_{3 \times 3}])$ & -- \\
  & $Q_0$      &
    \(
    \begin{array}{ll}
    \sigma_{CV} = \left\{ \begin{array}{l} 0.05G\sqrt{\Delta t} \\ 4.0G\sqrt{\Delta t} \end{array} \right. &
    \sigma_{CA} = \left\{ \begin{array}{l} 0.01G\sqrt{\Delta t} \\ 2.5G\sqrt{\Delta t} \end{array} \right. \\[2ex]
    \sigma_{SR} = \left\{ \begin{array}{l} 0.01G\sqrt{\Delta t} \\ 2.5G\sqrt{\Delta t} \end{array} \right. &
    \sigma_{OS} = \left\{ \begin{array}{l} 0.02G\sqrt{\Delta t} \\ 2.0G\sqrt{\Delta t} \end{array} \right.
    \end{array}
    \)
  & -- \\
  & $R_0$ & $\mathrm{diag}([10^2 \bm{I}_{3 \times 3},\ 5^2 \bm{I}_{3 \times 3},\ (\frac{\pi}{180})^2 \bm{I}_{3 \times 3},\ (\frac{\pi}{180})^2 \bm{I}_{3 \times 3}])$ & -- \\
  & $\alpha_{\text{AEKF}}$ & 0.92 & -- \\
  & $\alpha_{\text{SR}}$ & 0.05 & -- \\
  & $\omega_{\text{OS}}$ & 0.005 & -- \\
  & ${\Pi}$ &
  \(
  \begin{aligned}
  {\Pi} &= \begin{bmatrix}
  \pi_{11} & \pi_{12} & \cdots & \pi_{18} \\
  \pi_{21} & \pi_{22} & \cdots & \pi_{28} \\
  \vdots & \vdots & \ddots & \vdots \\
  \pi_{81} & \pi_{82} & \cdots & \pi_{88}
  \end{bmatrix}, \quad
  \pi_{ij} = \begin{cases}
  0.8 & i = j \\
  \frac{1}{35} & i \ne j
  \end{cases}
  \end{aligned}
  \)
  & -- \\
\midrule
\multirow{2}{*}{Seeker}
  & $\sigma_{FOV}$ & $15$ & deg \\
  & $\dot{\sigma}_{FOV}$ & $ 500 $ & deg/s \\
  & $R_{FOV}$ & $ 20 $ & km \\
\bottomrule\bottomrule
\end{tabular}
\label{table:4_1_parameters}
\end{adjustbox}
\end{table}
}


The simulation parameters summarized in Table \ref{table:4_1_parameters} establish the baseline configuration for evaluating the proposed guidance framework. A fixed missile integration step of 0.01 seconds ensures numerical accuracy, and the final MD threshold of 10 m provides a benchmark for successful interception. Within the MPCG, weighting matrices are designed to capture the relative significance of state variables, with $t_{horizon}=0.5$ seconds and $N=7$ collocation points empirically selected to balance accuracy and computational cost. The AEKF-IMM employs $P_0$ and $R_0$ to reflect state uncertainty and sensor noise, while $\alpha_{AEKF}$ and $\Pi$ enable adaptive covariance adjustment and robust model switching. Eight maneuver models, spanning CV, CA, Singer, and Oscillatory dynamics at low and high intensities, are incorporated. Finally, seeker constraints of a $15^\circ$ FOV limit on the allowable look angle, a $150^\circ/s$ limit on its rate, and 20 km detection range define the operational guidance envelope.

To validate the proposed framework, each scenario is simulated to assess both guidance and estimation performance. The evaluation metrics are organized into two groups:

\begin{itemize}
    \item \textbf{Guidance-related metrics} focus on trajectory behavior, compliance with seeker FOV limits, bounded missile acceleration under MPC constraints, final MD, and TTI performance.

    \item \textbf{Estimation-related metrics} include measurement and state estimation errors, as well as the consistency of model probability transitions across maneuver modes.
    
\end{itemize}

Overall, these metrics provide a comprehensive basis for verifying the feasibility and precision of the integrated MPCG and AEKF-IMM framework under maneuvering target conditions.

\subsection{Single Maneuver Scenarios}

Table \ref{table:4_2_single_maneuver_init} summarizes the initial conditions of the target and missile for three representative maneuvering scenarios: weaving in the pitch and yaw plane, and the barrel-roll maneuver. The target is initialized at a speed of 400 m/s with a maximum acceleration capability of $8G$, while the missile starts at 650 m/s with a maximum acceleration of $25G$. The simulation horizon is set to $t_f=50$ seconds. Additional parameters include the oscillation rate $\omega_{br}$ for weaving and barrel-roll maneuvers, the first-order lag constant $\tau_M$ for the missile autopilot, and the navigation constant $N_{ppng}$ used in the PPNG baseline law.

{\scriptsize
\begin{table}[!h]
\centering
\caption{Initial conditions of each single maneuver}
\begin{adjustbox}{max width=\columnwidth}
\begin{tabular}{llccccl}
    \toprule
    \textbf{Component} & \textbf{Parameter} & \textbf{Weaving (pitch)} & \textbf{Weaving (yaw)} & \textbf{Barrel-roll} & \textbf{Unit} \\
    \midrule
    \multirow{6}{*}{Target}
    & $\bm{r}_T(0)$      & [5,5,-5]   & [5,5,-5]   & [5,5,-5]   & km \\
    & $V_T(0)$           & 400       & 400       & 400       & m/s \\
    & $\gamma_T(0)$      & 0         & 0         & 45        & deg \\
    & $\chi_T(0)$        & 45        & 45        & 45        & deg \\
    & $V_{T,\max}$       & 550       & 550       & 550       & m/s \\
    & $a_{T,\max}$       & 8G        & 8G        & 8G        & m/s\textsuperscript{2} \\
    \midrule
    \multirow{6}{*}{Missile}
    & $\bm{r}_M(0)$      & [0,0,-5]   & [0,0,-5]   & [0,0,0]   & km \\
    & $V_M(0)$           & 650       & 650       & 650       & m/s \\
    & $\gamma_M(0)$      & 0         & 0         & 35.2644   & deg \\
    & $\chi_M(0)$        & 45        & 45        & 45        & deg \\
    & $V_{M,\max}$       & 800       & 800       & 800       & m/s \\
    & $a_{M,\max}$       & 25G       & 25G       & 25G       & m/s\textsuperscript{2} \\
    \midrule
    \multirow{5}{*}{}
    & $t_f$              & 50        & 50        & 50        & sec \\
    & $\omega_w$, $\omega_{br}$ & 0.25 & 0.25 & 0.25 & -- \\
    & $\tau_M$           & 0.05      & 0.05      & 0.05      & -- \\
    & $N_{ppng}$         & 3         & 3         & 3         & -- \\
    & $ K $              & 1         & 1         & 1         & -- \\
    \bottomrule
\end{tabular}
\label{table:4_2_single_maneuver_init}
\end{adjustbox}
\end{table}
}

\begin{figure}[!h]
  \centering
  \includegraphics[width=\columnwidth]{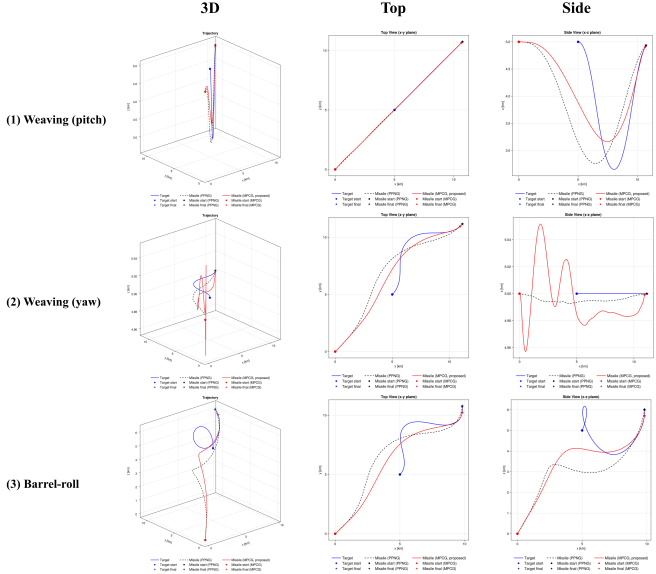}
  \caption{Trajectories of the missile and target for each scenario}
  \label{fig:4_2_trajectory}
\end{figure}

The resulting missile and target trajectories are illustrated in Fig. \ref{fig:4_2_trajectory}, where the target trajectory is shown in blue, the PPNG-guided missile in black dashed lines, and the MPCG-guided missile in red. The initial and final positions are marked with circles and stars, respectively. In both the pitch-weaving and barrel-roll cases, MPCG produced trajectories closely aligned with PPNG but achieved interception with a reduced TTI. In the yaw-weaving scenario, the MPCG trajectory exhibited larger oscillations relative to PPNG. However, despite the initial 5 km separation, the vertical deviations (5.04 km upward and 4.96 km downward) remained within acceptable bounds, demonstrating compliance with operational constraints.

Finally, Table \ref{table:4_3_single_eval} reports the quantitative performance metrics, specifically the final MD and TTI derived from the terminal positions in Fig. \ref{fig:4_2_trajectory}. The interception criterion is defined as the instant when the closing velocity becomes negative within the admissible intercept range, with the corresponding relative distance recorded as the final MD. All metrics are computed using a fourth-order Runge–Kutta (RK4) ODE solver for numerical accuracy.

\begin{table}[htbp]
\centering
\caption{Comparison of guidance performance for single-maneuver target scenarios}
\label{tab:guidance_comparison}
\begin{tabular}{llcc}
\toprule
\textbf{Target Maneuver} & \textbf{Guidance} & \textbf{Final MD [m]} & \textbf{TTI [sec]} \\
\midrule
\multirow{2}{*}{(1) Weaving (pitch)} 
& PPNG & 1.9335 & 23.95 \\
& \textbf{MPCG} & \textcolor{red}{\textbf{1.6662}} & \textbf{23.66} \\
\midrule
\multirow{2}{*}{(2) Weaving (yaw)} 
& PPNG & 1.1121 & 25.45 \\
& \textbf{MPCG} & 1.9983 & \textcolor{red}{\textbf{24.72}} \\
\midrule
\multirow{2}{*}{(3) Barrel-roll} 
& PPNG & 1.0845 & 29.32 \\
& \textbf{MPCG} & \textbf{0.9761} & \textcolor{red}{\textbf{27.81}} \\
\bottomrule
\end{tabular}
\label{table:4_3_single_eval}
\end{table}

\begin{figure*}[!t]
  \centering
  \includegraphics[width=\textwidth]{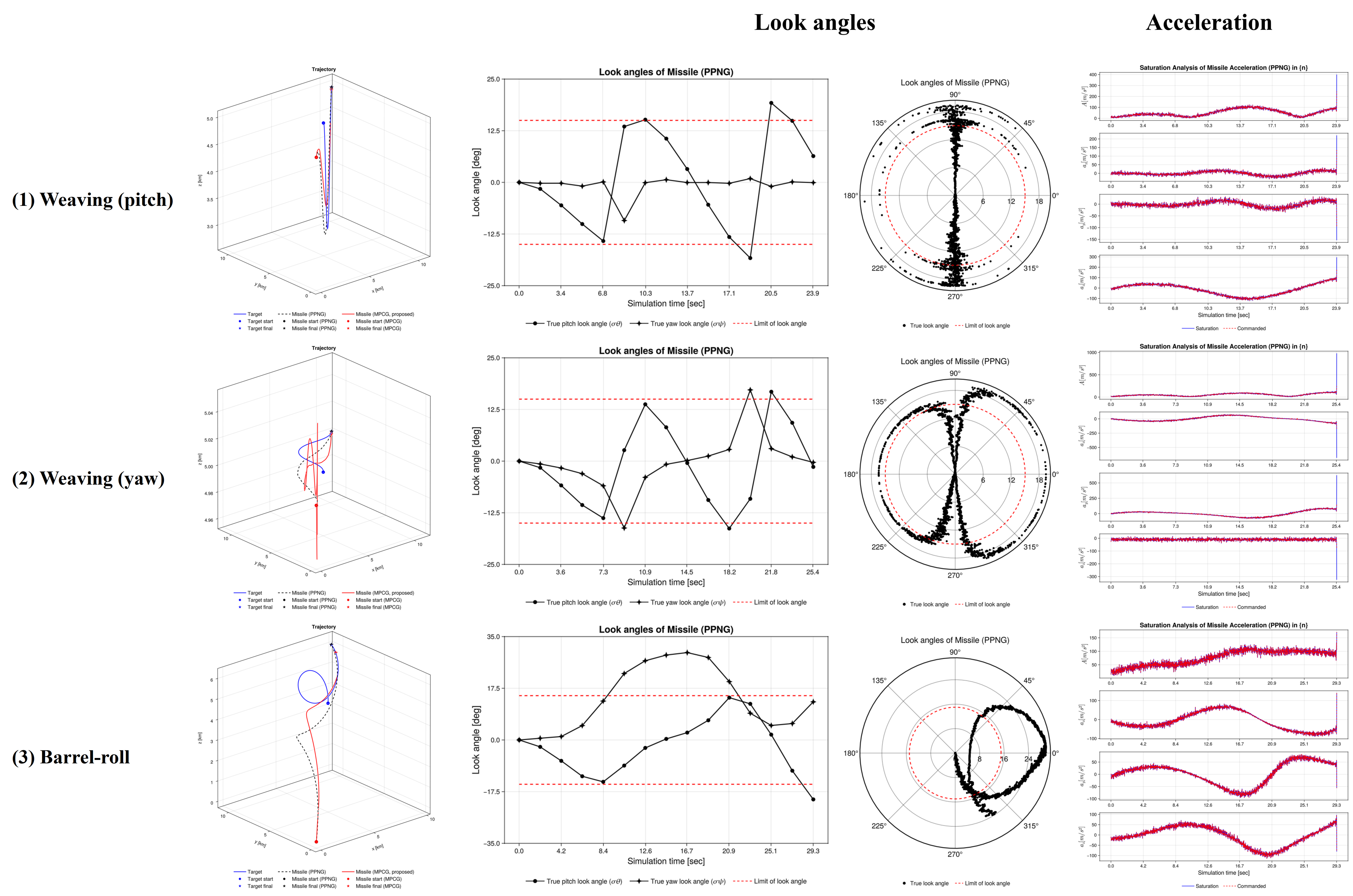}
  \caption{Constraint compliance of look angles and missile acceleration in PPNG scenarios}
  \label{fig:4_4_ppng_result_1}
\end{figure*}

The comparative results for each scenario are as follows:
In the weaving in pitch case, the TTI achieved by MPCG was 23.66 seconds, which is approximately 0.3 seconds faster than PPNG. The final MD was also reduced by approximately 0.267 meters in MPCG, indicating enhanced interception accuracy.

In the weaving in yaw scenario, although MPCG did not outperform PPNG in terms of MD, it still provided a reasonable performance, with a TTI about 0.73 seconds shorter than that of PPNG.

In the barrel-roll scenario, MPCG showed better performance in both MD and TTI compared to PPNG.
In summary, across the three representative maneuver scenarios, MPCG demonstrated either comparable or improved performance over PPNG in both MD and TTI, thereby validating the effectiveness of the proposed guidance scheme.

Fig. \ref{fig:4_4_ppng_result_1} illustrates the look angle and missile acceleration constraints under PPNG guidance across the three engagement scenarios.
In all cases, the look angles exceed the allowable FOV, both in the time-domain plots and in the phase space ($\theta-\psi$ plane), as indicated by the red dashed lines or circular boundaries.

Additionally, the missile acceleration plots show that as the engagement progresses toward interception, the guidance commands approach or exceed the saturation limits (blue solid line), particularly along the lateral axes. These findings highlight a critical limitation of PPNG in seeker-constrained environments, motivating the need for constraint-aware guidance strategies.

\begin{figure*}[!h]
  \centering
  \includegraphics[width=0.72\textwidth]{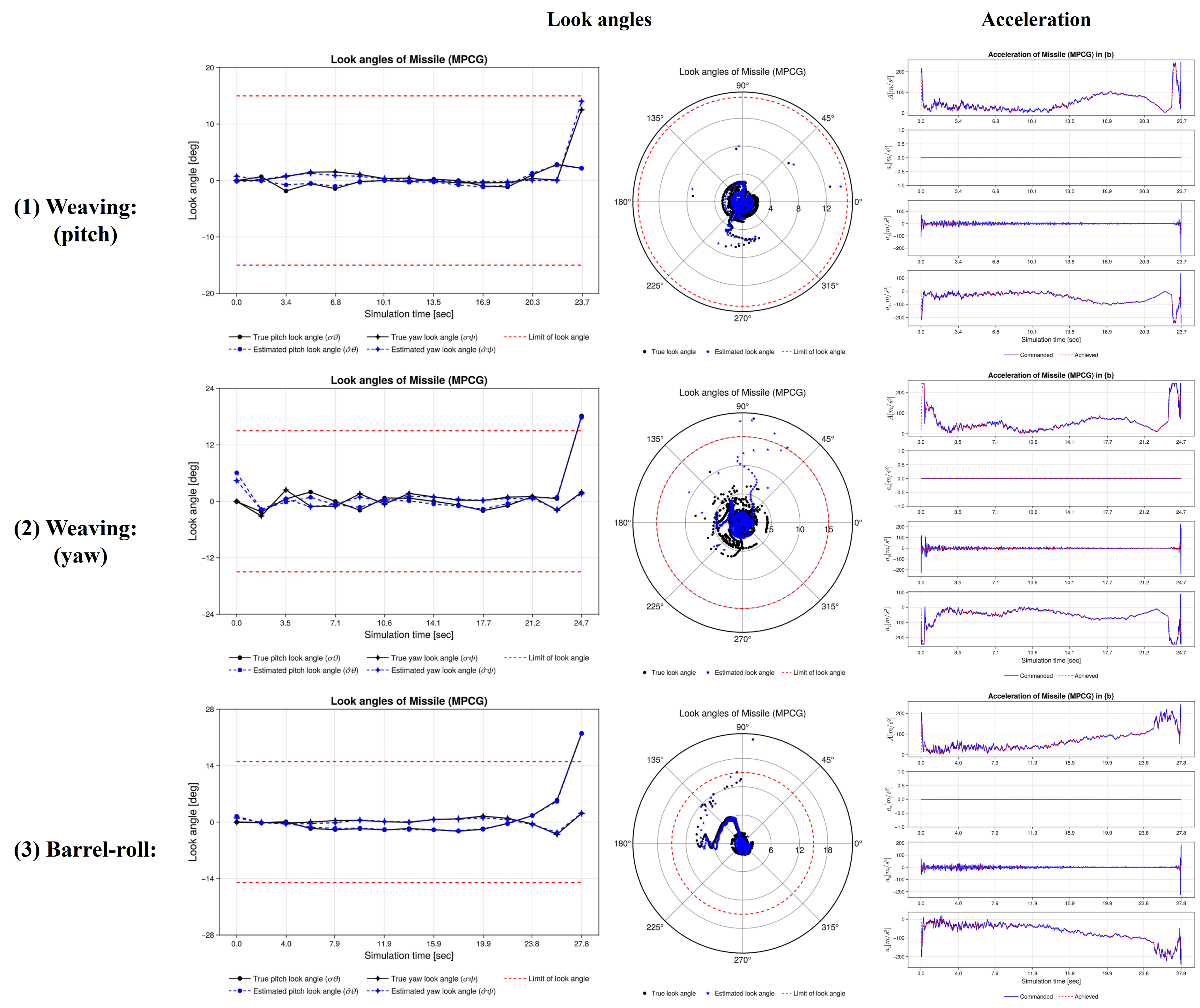}
  \caption{Constraint compliance of look angles and missile acceleration in MPCG scenarios}
  \label{fig:4_5_mpcg_result_1}
  \vspace{5.0ex}
  \centering
  \includegraphics[width=0.74\textwidth]{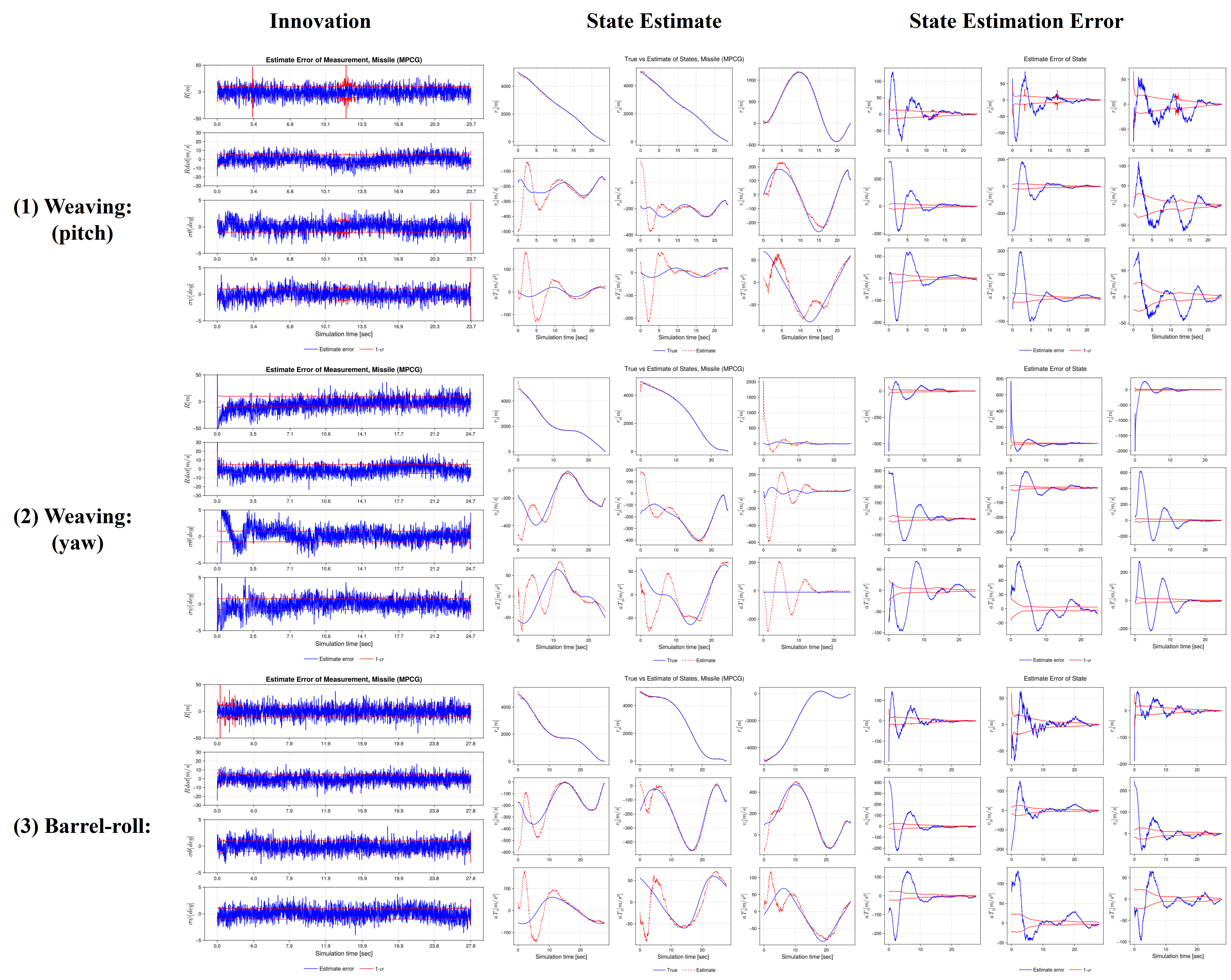}
  \caption{Measurement and state estimation accuracy of the AEKF in MPCG scenarios}
  \label{fig:4_6_mpcg_result_2}
\end{figure*}

Fig. \ref{fig:4_5_mpcg_result_1} depicts the behavior of the missile's look angles and acceleration commands under the MPCG scheme across three engagement scenarios.

The look angle results show the look angle trajectories over time and in the azimuth-elevation phase domain, respectively. Unlike the results of the PPNG guidance shown earlier, the MPCG maintains the missile’s look angle consistently within the allowable FOV limits throughout most of the engagement. In both the weaving-in-yaw and barrel-roll scenarios, slight violations are observed immediately prior to interception. This minor exceedance is attributed to the aggressive terminal maneuvering required to reduce the final MD while operating close to the edge of the engagement envelope, where the seeker’s FOV constraint becomes difficult to maintain due to the sharp look angle dynamics.

The acceleration results present the missile acceleration magnitude and its components in the missile’s body-fixed frame. The solid blue lines represent the acceleration commands generated by MPCG, which remain safely within the maximum allowable acceleration limit of 25G. Moreover, the first-order lag autopilot dynamics ensure smooth responses by attenuating sudden spikes in the command signals, resulting in practical and feasible missile control actions.

These results demonstrate the effectiveness of MPCG in satisfying both look angle and acceleration constraints, thereby ensuring reliable and physically realizable guidance performance across diverse maneuvering scenarios.

Fig. \ref{fig:4_6_mpcg_result_2} presents the estimation results of the AEKF used within the MPCG framework. The results are organized into three columns showing measurement estimates error (innovation), estimated state trajectories, and corresponding state estimation errors.

In the innovation results, the measurement estimation errors are shown, and most values lie within the $1-\sigma$ covariance bounds, indicating reliable estimation performance. In Kalman filter design, the measurement estimation error is often used as a key criterion to evaluate the filter’s effectiveness. Both $1\sigma$ and $3\sigma$ bounds are commonly used to assess estimation quality; this study adopts the $1\sigma$ criterion as the threshold for acceptable accuracy.

The state estimation results include the position, velocity, and acceleration estimated by the AEKF, while the estimation errors represent the differences between the true and estimated states. Particular attention is given to the acceleration estimates in each scenario, as they directly reflect the accuracy of target acceleration estimation.

Initially, due to large initial covariance values, the acceleration estimates exhibit noticeable discrepancies. However, as the simulation progresses, the estimation errors decrease, as shown in the state estimation errors. This trend confirms the filter's increasing accuracy over time, ultimately supporting improved prediction performance in the MPCG scheme.

Furthermore, the adaptive structure of the AEKF contributed to mitigating filter divergence by dynamically adjusting the process and measurement noise covariances for each model, ensuring stable estimation throughout the maneuver.

\begin{figure}[!htbp]
  \centering
  \includegraphics[width=\columnwidth]{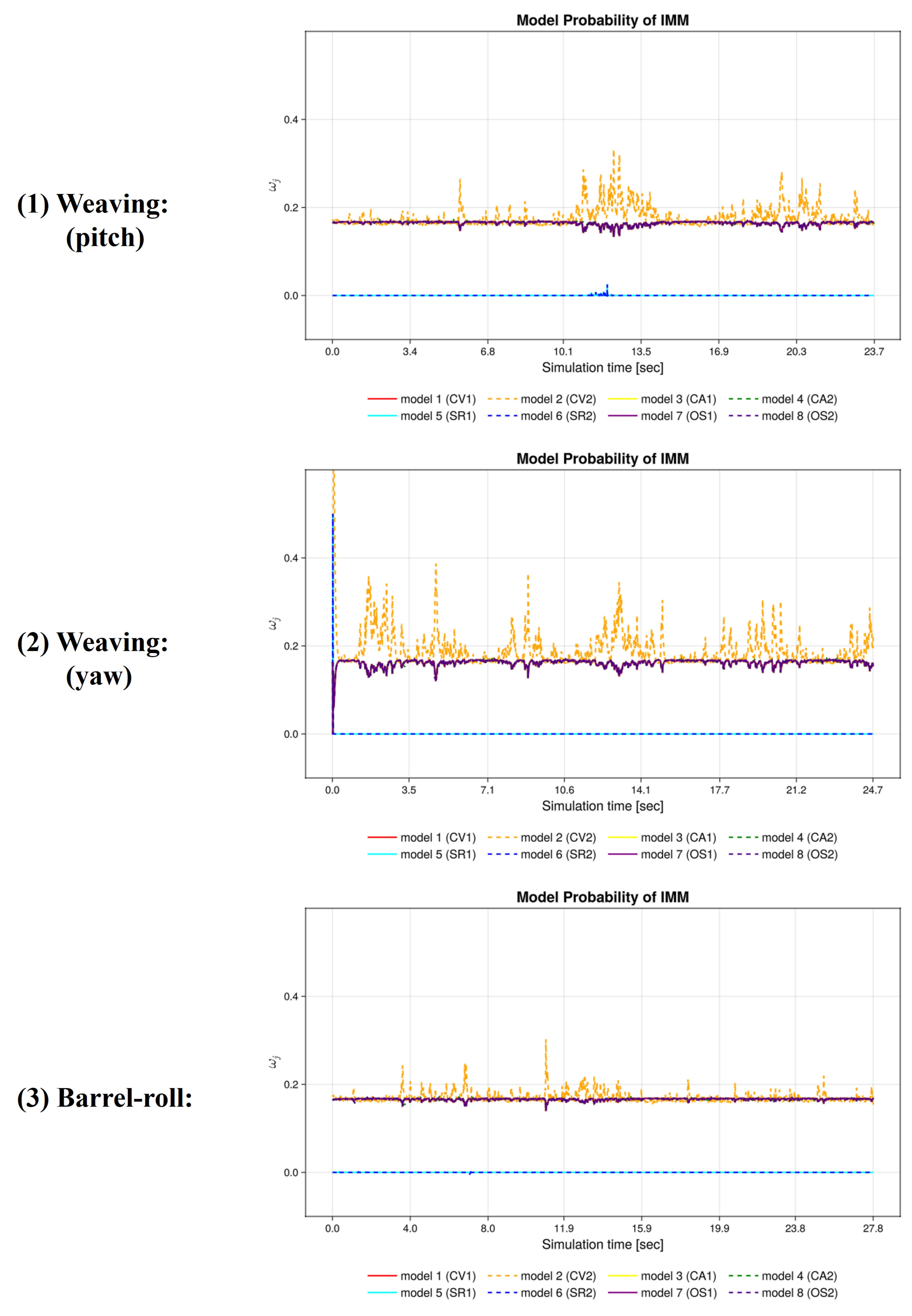}
  \caption{Model probabilities of the IMM for each scenario}
  \label{fig:4_7_mpcg_result_3}
\end{figure}

Fig. \ref{fig:4_7_mpcg_result_3} presents the time history of the model probabilities, denoted as $\omega_j$, generated by the IMM estimator for three representative target maneuver scenarios.

The results show the time evolution of the model probabilities for each of the eight maneuvering models in the IMM bank. These consist of low- and high-intensity versions of CV, CA, Singer, and Oscillatory models.

Each line in the plots represents the likelihood of a specific model being active at a given time; values close to one indicate that the model most accurately describes the current target maneuver.

Across all three scenarios, Model 2 (CV2)—the high-intensity version of CV—exhibited clearly dominant probability levels particularly during periods of strong curved motion of the target. This indicates that CV2 aligned most consistently with the actual maneuver characteristics during those times.

In contrast, Singer models generally maintained low probability values and did not effectively capture the dominant maneuvering patterns. Occasional increases in the probability of other models were observed during transient phases or at maneuver initiation points, which can be interpreted as temporary fluctuations due to model switching or initial estimation uncertainty.

These findings highlight the capability of the IMM structure to operate multiple maneuvering models concurrently and identify the most representative target motion profile in online, even in complex 3D engagement scenarios. The results confirm that the IMM estimator adapts to changes in maneuvering characteristics and supports accurate input estimation for guidance algorithms like MPCG.


\section{Conclusions} \label{sec:5_conclusion}

This study presents an MPCG framework specifically designed for missiles equipped with strapdown seekers. The proposed guidance system integrates two major components: an LGRPM-based MPC module and an adaptive estimation framework using AEKF-IMM.

The MPC module is formulated by transcribing a nonlinear OCP into a NLP problem using the LGRPM. This formulation is based on look angle dynamics, making it structurally compatible with strapdown seekers that lack direct LOS rate measurements. Furthermore, the MPC includes explicit constraints for seeker FOV and missile acceleration limits to ensure physically feasible and sensor-compliant guidance commands.

To predict states at the collocation points within the MPCG framework, target acceleration estimates are required. These estimates are provided by an AEKF-IMM estimator, which is designed to adaptively handle varying target maneuvers.

The estimation framework employs an AEKF to estimate target acceleration online by tuning process and measurement noise covariances. This AEKF is integrated with an IMM architecture that includes eight maneuvering models—covering different levels of intensity for CV, CA, Singer, and Oscillatory—to handle a wide spectrum of target motion uncertainties.

To evaluate the performance of the proposed MPCG framework, three engagement scenarios—pitch and yaw weaving, barrel-roll—were simulated. Performance was assessed using metrics such as final MD, TTI, seeker FOV compliance, and missile acceleration limits. Overall, MPCG successfully guided the missile to interception under all scenarios while satisfying practical constraints, particularly showing effective terminal control and stable acceleration commands within 25G through a first-order lag autopilot.

However, the results also reveal realistic limitations. During aggressive endgame maneuvers, especially in yaw weaving and barrel-roll cases, the look angle briefly exceeded the FOV just before interception. Additionally, although the AEKF-IMM estimator generally tracked target acceleration well, transient estimation errors occurred at maneuver transition points, which later recovered within the $1-\sigma$ bound. These findings highlight that while MPCG offers a constraint-aware and effective solution for strapdown seeker systems, performance during target maneuver transitions requires careful modeling and estimation strategies.

\bibliographystyle{IEEEtran}
\bibliography{ref}

\end{document}